\documentclass[sigconf]{acmart}

\usepackage{amsfonts,amsthm,listings,fancyhdr,enumerate,enumitem,subcaption,euscript,fixmath}

\newcommand{\snipbegin}[1]{}
\newcommand{\snipend}[1]{}

\newcommand{\reals}{\mathbb{R}}

\newcommand{\mesh}{\mathbb{M}}
\newcommand{\vertices}{\mathbb{V}}
\newcommand{\edges}{\mathbb{E}}
\newcommand{\terrain}{\Sigma}

\newcommand{\varregion}{R}

\newcommand{\filltime}{\tau}
\newcommand{\initrain}{R}
\newcommand{\localflow}{\lambda}

\newcommand{\edgecross}{\mathbold{E}}
\newcommand{\edgeuniquecross}{\mathbold{\hat{E}}}
\newcommand{\flowrate}{\phi}
\newcommand{\flowgraph}{\EuScript{M}}

\newcommand{\varheight}{\chi}

\newcommand{\fillrate}{F}
\newcommand{\spillrate}{S}

\newcommand{\paths}{\mathbb{P}}
\newcommand{\profile}{\EuScript{P}}
\newcommand{\faces}{\EuScript{F}}

\newcommand{\mergetree}{\mathsf{T}}

\newcommand{\tribdag}{\mathsf{G}}

\newcommand{\volume}{\text{Vol}}  
\newcommand{\BigO}{O}  
\newcommand{\BigTheta}{\Theta}  
\newcommand{\varvolume}{\psi}

\newcommand{\varsinks}{\rho}

\newcommand{\saddle}{\text{Sd}}
\newcommand{\sinks}{\text{Sk}}
\newcommand{\tribs}{b}

\newcommand{\channel}{\EuScript{C}}

\newcommand{\Scan}{\operatorname{Scan}}
\newcommand{\Sort}{\operatorname{Sort}}
\newcommand{\varmem}{m}
\newcommand{\varblock}{b}

\newcommand{\lb}{\mathrm{lb}}
\newcommand{\rb}{\mathrm{rb}}

\newcommand{\etal}{\textit{et al}.}

\newcommand{\header}[1]{\smallskip\noindent\textbf{#1}}

\newcommand{\rsec}[1]{Section~\ref{sec:#1}}

\theoremstyle{definition}
\newtheorem{thm}{Theorem}[section]
\newtheorem{lemma}[thm]{Lemma}

\begin{document}

\setcopyright{acmlicensed}
\author{Aaron Lowe}
\affiliation{%
  \institution{Duke University}
}

\author{Svend C.\@ Svendsen}
\affiliation{%
  \institution{Aarhus University}
}

\author{Pankaj K.\@ Agarwal}
\affiliation{%
  \institution{Duke University}
}

\author{Lars Arge}
\affiliation{%
  \institution{Aarhus University}
}

\title{1D and 2D Flow Routing on a Terrain}
\copyrightyear{2020}
\acmYear{2020}
\setcopyright{acmcopyright}\acmConference[SIGSPATIAL '20]{28th International Conference on Advances in Geographic Information Systems}{November 3--6, 2020}{Seattle, WA, USA}
\acmBooktitle{28th International Conference on Advances in Geographic Information Systems (SIGSPATIAL '20), November 3--6, 2020, Seattle, WA, USA}
\acmPrice{15.00}
\acmDOI{10.1145/3397536.3422269}
\acmISBN{978-1-4503-8019-5/20/11}
%
%

\date{\today}

\begin{abstract}
An important problem in terrain analysis is modeling how water flows
across a terrain  creating floods by forming channels and filling depressions. 
In this paper we study a number of \emph{flow-query} related problems:
Given a terrain $\terrain$, represented as 
a triangulated $xy$-monotone surface with $n$ vertices,
a rain distribution $\varregion$ which may vary over time,  
determine how much water is flowing over a given edge as a function of time. 
We develop internal-memory as well as I/O-efficient algorithms for flow queries.

This paper contains four main results:

    (i) We present an internal-memory algorithm that preprocesses $\terrain$ into a linear-size data structure 
    in $\BigO(n \log n)$ time that for a (possibly time varying) rain distribution $\varregion$ can return the flow-rate functions of all edges of $\terrain$ in $\BigO(\varsinks k+|\flowrate| \log n)$ time, where $\rho$ is the number of sinks in $\Sigma$, $k$ is the number of times the rain distribution changes, and 
    $|\flowrate|$ is the total complexity of the flow-rate functions that have non-zero values;
    $|\flowrate| = \Theta(n (\varheight +k))$ in the worst case, where $\varheight$ is the height of the merge tree of $\terrain$ and $k$ is the number of times the rain distribution changes, but $|\flowrate|$ is much smaller in practice. 

    (ii) We also present an I/O-efficient algorithm for preprocessing $\terrain$ into a linear-size data structure using $\BigO(\Sort(n))$ I/Os and $\BigO(n \log n)$ internal computation time, 
    so that for a rain distribution $\varregion$, it can compute the flow-rate function of all edges 
    using $\BigO(\text{Sort}(|\flowrate|))$ I/Os and $\BigO(|\flowrate| \log |\flowrate|)$ internal computation time. 

    (iii) $\terrain$ can be preprocessed in $\BigO(n \log n)$ time, into a linear-size data structure 
    so that for a given rain distribution $\varregion$, 
    the flow-rate function of an edge $(q,r) \in \terrain$ under the single-flow direction (SFD) model can be computed in $\BigO(|\varregion| +|\tribs_q| k\log n)$ time, where $|\varregion|$ is the number of vertices in $\varregion$ on which nonzero rain falls,
    and $|\tribs_q|$ is the number of tributaries of $q$ in which rain directly falls in. 

    (iv) We present an algorithm for computing the two-dimensional channel along which water flows using Manning's equation; a widely used empirical equation that relates the flow-rate of water in an open channel to the geometry of the channel along with the height of water in the channel. 
    Given the flow-rates along a path of edges, the algorithm computes the height of water and boundary along the channel in $\BigO(|\channel| \log |\channel|)$ time, where $|\channel|$ is the number of wetted faces at least partially covered by water in the channel. 
\end{abstract}

\maketitle

\section{Introduction}
\label{sec:intro}

An important problem in terrain analysis is modeling how water flows
across a terrain and creates floods by forming channels and filling up depressions.
The rate at which a depression fills up during a rainfall depends not only on the shape of the depression  
and the size of its watershed (i.e.,  the area of the terrain that contributes water to the depression)  
but also on other depressions that are filling up.
Water falling on the watershed of a filled depression flows to a neighboring depression
effectively making the watershed of the latter larger and filling it up faster.
Modeling how depressions fill and how water spills into other depressions
during a flash flood event is therefore an important computational problem.

Besides determining which areas of a terrain become flooded and when they become flooded, 
determining the 2D channels (rivers) along which water flows across the terrain is also important. 
The flow queries we ask can also be used to solve related flood-risk queries, 
and the algorithms developed will provide a simpler and faster algorithm for a previously studied flood-risk queries. 
We assume we are given a terrain $\terrain$, represented as a triangulated $xy$-monotone surface with $n$ vertices. 
As in earlier papers,~\cite{RLA17,lowe2019flood,ARZ10}
we assume that water flows along the edges of $\terrain$. 
Two models of water flow along edges have been proposed:
(i) a simple and more widely used model called the single flow-direction (SFD) model in which 
water from a vertex flows along \textit{one} of its downward edges, and 
(ii) a more accurate but more complex model called the multi-flow-direction (MFD) model in which water
at a vertex splits and flows along all of its downward edges. There is also some earlier work by Liu and Snoeyink~\cite{LS05} which allows water to flow in the interior of edges. 
We consider both of these models.

We study the following three problems in this paper:
\begin{itemize}
    \item \textit{Terrain-flow query}: given a rain distribution (possibly varying with time), compute as a function of time the flow rate (of water) for all edges of $\terrain$. 
    \item \textit{Edge-flow query}: given a rain distribution and a query edge $e$ of $\terrain$, compute the flow rate of $e$ under the single flow-direction (SFD) model. 
    \item \textit{2D flow network}: Given a 1D flow network, represented as a set of edges along with their flow values, compute 2D channels along which water flows. 
\end{itemize}

Finally, as high-resolution terrain data sets are becoming easily available, 
their size easily exceeds the size of main memory of a standard computer, 
so movement of data between main memory and external memory (such as disk) becomes the bottleneck in computations. 

We use the \textit{I/O-model} with one disk by Aggarwal and Vitter~\cite{AV88}, 
in which, the computer is equipped with a two-level memory hierarchy
consisting of an \textit{internal memory}, which is capable of holding $\varmem$ data items,
and an \textit{external memory} of unlimited size.
All computation happens on data in internal memory.
Data is transferred between internal and external memory in blocks of $\varblock$
consecutive data items.
Such a transfer is referred to as an \textit{I/O-operation} or an \textit{I/O}.
The cost of an algorithm is the number of I/Os it performs.
The number of I/Os required to read $n$ items from disk is $\Scan(n) =
\BigO(n/\varblock)$.
The number of I/Os required to sort $n$ items is $\Sort(n) =
\BigTheta \big((n/\varblock) \log_{\varmem/\varblock}(n/\varblock)\big)$~\cite{AV88}.
For all realistic values of $n$, $\varmem$, and $\varblock$ we have $\Scan(n) <
\Sort(n) \ll n$.

\header{Related work.}
Due to its importance, the problem of modeling how water flows on a terrain
has been studied extensively, and many approaches have been
taken to address this problem.
One approach focuses on accurately simulating fluid dynamics using non-linear
partial differential equations such as the Navier-Stokes equations.
These equations have no closed form solutions and are usually solved using numerical methods.
They often account for additional factors, such as the effects of different
terrain types and drainage networks.
While these models tend to be more accurate, they are
computationally expensive and do not scale to large terrains.
Bates and De Roo \cite{bates2000simple} developed one such model for simulating flooding on digital elevation models (DEMs) using two flow models for different regions of the terrain: the first handles flow within rivers and the second models flow of water as it spreads over floodplains.
While there has been some research into refining the representation of channels, such as Wood \etal~\cite{wood2016calibration}, often the channel geometry is assumed to be a simple model (e.g. rectangular or trapezoidal.)

Water-flow modeling on a terrain also has been studied in the
GIS community.
These approaches use simpler models, which are computationally efficient and
suitable for large datasets. 
Although some early work, e.g.~\cite{LS05} 
allowed water to flow in the interior of faces, recent work assumes that water flows 
along the edges of the terrain and the SFD and MFD models described above are used to model 
the water flow locally at a vertex, see e.g. ~\cite{RLA17,lowe2019flood, ARZ10}
%
Arge et al.~\cite{arge2000grid} described an I/O-efficient algorithm  for 
the \textit{flow-accumulation} problem in the SFD model,
water falls uniformly on the terrain which asks 
how much water flows over each point in a terrain 
assuming the terrain has only one sink at infinity. 
Their algorithm performs a total of $\BigO(\Sort(n))$ I/Os, where $\Sort(n)$
is the number of I/Os required to sort $n$ items.
The flow accumulation model only provides rough solution to flow
modeling, since it assumes that either the terrain does not have any local minima 
or that they have been filled in advance. 

In order to accurately model flow it is necessary to compute times at which
depressions fill and simulate how water spills from one depression into
others.
Arge et al.~\cite{ARZ10} proposed the first I/O-efficient algorithm 
that computes the fill times of all maximal depressions in
$\BigO(\Sort(\varsinks) \log(\varsinks/\varmem))$ I/Os, where $\varsinks$ is
the number of depressions in the terrain and $\varmem$ is the size of the
internal memory.
If $\varsinks = \BigO(\varmem)$, the algorithm can be simplified
and requires only $\BigO(\Sort(n))$ I/Os.

Arge et al.~\cite{ARRR17} described an I/O-efficient algorithm that computes
which points on the terrain become flooded if a total volume $\varvolume$ of rain falls
according to a distribution $\varregion$.
It performs $\BigO(\Sort(n) + \Scan(\varheight \cdot \varsinks))$
I/Os, where $\varheight$ is the height of the merge tree of the terrain.
In the worst case $\varheight \varsinks = \Omega(n^2)$, 
 but it can be bounded to $\BigO(\Sort(n))$ under certain practical assumptions.

Lowe et al.~\cite{lowe2019flood} presented efficient algorithms for several flood queries on a terrain under the \textit{multiflow direction} (MFD) model.
They presented a $\BigO(n \log n)$-time algorithm to answer terrain flood queries.
They also showed that 
a terrain $\terrain$ can be preprocessed in $\BigO(n \log n + n \varsinks)$-time
into a data structure that can answer \textit{point-flood} queries:
Given a rain distribution $\varregion$, a volume of rain $\varvolume$, and a
point $q \in \terrain$, determine whether $q$ will be flooded.
The query time is $\BigO(|\varregion|\tribs_q +
\tribs_q^2)$ time, where $\tribs_q$ is the number of maximal depressions that contain the
query point $q$;
$\tribs_q = \Omega(n)$ in the worst case, but in practice it is much smaller. 
Finally, they presented a $\BigO(n\tribs_q + \tribs_{q}^\omega)$-time algorithm to 
determine when a query point $q$ gets flooded, 
where $\omega$ is the exponent of fast matrix multiplication. 
To our knowledge, no I/O-efficient algorithms are known for these flooding queries under the MFD model.

\header{Our results.}
There are four main results in the paper.

    (i) We present a $\BigO(n \log n)$ time algorithm for preprocessing $\terrain$ into a linear-size data structure for answering terrain-flow queries: for a rain distribution $\varregion$, it can compute the flow rate of all edges in $\BigO(\varsinks k+|\flowrate| \log |\flowrate|)$ time, where $|\flowrate|$ is total complexity of nonzero flow-rate functions, $\varsinks$ is the number of sinks in $\terrain$, and $k$ is the number of times the rain distribution changes. 
    In the worst case $|\flowrate| = \Theta(n (\varheight +k))$, where $\varheight$, as above, is the height of the merge tree of $\terrain$, but $|\flowrate|$ is much smaller in practice. An immediate corollary of our result is that a flood-time query (i.e. given $\varregion$ and a point $q \in \terrain$, when will $q$ be flooded) can be answered in the same time, which is a significant improvement over the result in \cite{lowe2019flood}. 
        \smallskip

(ii) We present two I/O-efficient algorithms for the terrain flow algorithm. 
  We first preprocess the terrain using $\BigO(\Sort(n))$ I/Os and $\BigO(n \log n)$ internal computation time. 
  The first algorithm that $\varsinks (\varheight + k) = \BigO(\varmem)$, where $\varheight$,
  $k$ and $\varsinks$ are as above, and considers a terrain-flow query in 
  $\BigO(\Sort(|\flowrate|))$ I/Os and $\BigO(|\flowrate| \log |\flowrate|)$ internal computation time.
  The second algorithm assumes $\varsinks = \BigO(\varmem)$ and answers a
  terrain-flow query in $\BigO(\Sort((\varheight + k) n \log n))$ I/Os and
  $\BigO((\varheight + k) n \log^2 (k n))$ internal computation time.
  We additionally note that since the terrain-flow query is a more general
  problem, these algorithms also yield I/O-efficient algorithms for the
  terrain-flood and flood-time queries under the MFD model studied in~\cite{lowe2019flood}.
        \smallskip

(iii) The terrain-flow query algorithm naturally can also be used to perform edge-flow queries. 
    Under SFD flow, we can build a linear sized data structure in $\BigO(n \log n)$ time 
    which given an edge $e = (q,r)$ and rain distribution $\varregion$ computes $\flowrate_{e}$ in $\BigO(|\varregion|+\tribs_{q}k \log n)$ time, where $|\varregion|$ is the complexity of the rain distribution, $\tribs_{q}$ is the number of tributaries of $q$ in which rain is falling, and $k$ is the number of times the rain distribution changes.
        \smallskip

(iv) We present an algorithm that given a 1D flow network, represented as a path of edges along with their flow values, 
    identifies the depth and width of water along the 2D flow network in $\BigO(|\channel| \log |\channel|)$ time, where $|\channel|$ is the number of ``wetted'' faces at least partially covered by the water in the channel. 
    We do so by utilizing Manning's equation \cite{manning1890flow}, a widely used empirical formula relating flow-rate of water in an open channel to the geometry of the channel. 
    The algorithm utilizes the key idea that while the profile of the channel changes continuously as we sweep along the 1D flow network, the combinatorial structure only changes at discrete events. 
    By tracking when these events occur, the algorithm efficiently computes the depth of the river along the flow network. 
    We note that previous work computing the 2D channel assumes the cross section of each channel has a simple geometry (e.g. rectangular or trapezoidal) \cite{bates2000simple}. 
    In contrast, we do not make any such assumption.

We note that our terrain-flow query algorithm builds on some ideas in \cite{lowe2019flood} namely, 
it also performed a sweep in the $(-z)$-direction. 
However, our new algorithm is more general, simpler, faster, and several new ideas are needed.
First note that the terrain-flood and flood-query problems studied in \cite{lowe2019flood} are special cases of the terrain-flow query problem. 
In the process of computing the flow rate of all edges, it computes the flood-time of all depressions of $\terrain$. 
In contrast, the algorithms in \cite{lowe2019flood} either compute which depression got flooded or the flood time of only one depression. 
Furthermore, computation of the latter required a complicated algorithm that relied on matrix multiplication and ran in $\BigO(n^{\omega})$ time in the worst case, while our algorithm takes $\BigO(n^2 \log n)$ time in the worst case. 
Additionally, our algorithm can handle rain distributions that vary over time, and our algorithm is output-sensitive. 
Finally, we also present an I/O-efficient algorithm for answering terrain-flow queries.

\section{Preliminaries \& Models}
\label{sec:preliminaries}

In this section we give a number of preliminary definitions and describe the flooding model, most of the text here follows closely ~\cite{RLA17,lowe2019flood}.

\subsection{Geometric Preliminaries}

\snipbegin{AMZ15}
\header{Terrains.}
Let $\mesh$ be a triangulation of $\reals^2$,
and let $\vertices$ be the set of vertices of $\mesh$;
set $n = |\vertices|$. We assume that $\vertices$ contains
a vertex $v_\infty$ at infinity, and that each edge $\{u, v_\infty\}$ is a
ray emanating from $u$; the triangles in $\mesh$ incident to $v_\infty$
are unbounded. Let $h : \mesh \to \reals$ be a height function. We
assume that the restriction of $h$ to each triangle of $\mesh$ is
a linear map, that $h$ approaches $+\infty$ at $v_\infty$, and that the
heights of all vertices are distinct. Given $\mesh$ and $h$, the graph
of $h$, called a \emph{terrain} and denoted by $\terrain = (\mesh, h)$, is an $x y$-monotone
triangulated surface whose triangulation is induced by $\mesh$.
\snipend{AMZ15}

\header{Critical vertices.}
There is a natural cyclic order on the neighbor vertices of a vertex $v$ of $\mesh$, 
and each such vertex is either an \emph{upslope} or \emph{downslope} neighbor.
If $v$ has no downslope (resp.\@ upslope) neighbor,
then $v$ is a \emph{minimum} (resp.\@ \emph{maximum}).
We also refer to a minimum as a \emph{sink}.
If $v$ has four neighbors $w_1$, $w_2$, $w_3$, $w_4$ in clockwise order
such that $\max(h(w_1), h(w_3)) < h(v) < \min(h(w_2), h(w_4))$
then $v$ is a \emph{saddle} vertex.

\header{Level sets, contours, depressions.}
Given $\ell \in \reals$,
the \emph{$\ell$-sublevel set of $h$}
is the set $h_{< \ell} = \{ x \in \reals^2 \mid h(x) < \ell \}$,
and the \emph{$\ell$-level set of $h$}
is the set $h_{= \ell} = \{ x \in \reals^2 \mid h(x) = \ell \}$.
Each connected component of $h_{< \ell}$ is called a \emph{depression},
and each connected component of $h_{= \ell}$ is called a \emph{contour}.
Note that a depression is not necessarily simply connected,
as a maximum can cause a hole to appear; 

For a point $x \in \mesh$, a depression $\beta_x$ of $h_{< \ell}$ is said to be
\emph{delimited by the point $x$} if $x$ lies on the boundary of $\beta$,
which implies that $h(x) = \ell$.
\snipbegin{ARRR16}%
A depression $\beta_1$ is \emph{maximal} if every depression
$\beta_2 \supset \beta_1$ contains strictly more sinks than $\beta_1$.
A maximal depression that contains exactly one sink is called
an \emph{elementary depression}.
Note that each maximal depression is delimited by a saddle,
and a saddle that delimits more than one maximal depression
is called a \emph{negative saddle}. 
For a maximal depression $\beta$, let $\saddle(\beta)$ denote the saddle delimiting $\beta$, 
and let $\sinks(\beta)$ denote the set of sinks in $\beta$. 
The \emph{volume} of a depression $\beta$ of $h_{< \ell}$ is
\begin{equation}
  \volume (\beta) = \int_{\beta} (\ell - h(v)) dv.
  \label{eqn:volume}
\end{equation}
\begin{figure}[t]%
  \includegraphics{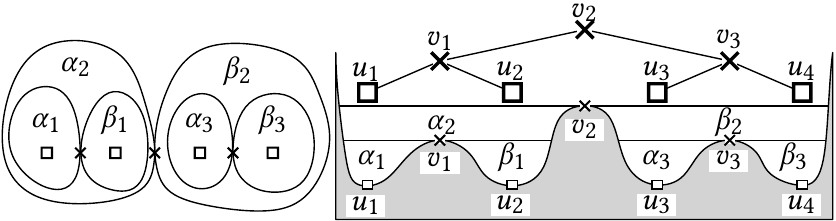}

  \begin{minipage}{0.4\linewidth}%
    \centering
    \subcaption{%
      \label{fig:terraintop}%
    }%
  \end{minipage}%
  \begin{minipage}{0.6\linewidth}%
    \centering
    \subcaption{%
      \label{fig:terrainside}\label{fig:terrainmt}%
    }%
  \end{minipage}%

  \caption{%
    \normalfont
    An example terrain with saddle vertices $v_1$-$v_3$.
    Each saddle $v_i$ delimits two maximal depressions $\alpha_i$ and $\beta_i$.
    \protect(\subref{fig:terraintop})~Terrain seen from above.
    Sinks are marked with a square and saddles are marked with a cross.
    \protect(\subref{fig:terrainside})~Terrain seen from the side,
  }
  \label{fig:terrain}%
\end{figure}

\header{Merge tree.}
The maximal depressions of a terrain form a hierarchy that is easily
represented using a rooted tree, called the \emph{merge tree} \cite{KOBPS97,CSA03}
and denoted by $\mergetree$.
Suppose we sweep a horizontal plane from $-\infty$ to $\infty$.
As we vary $\ell$, the depressions in $h_{< \ell}$
vary continuously, but their structure changes only at sinks and negative saddles.
If we increase $\ell$, then a new depression appears at a sink,
and two depressions merge at a negative saddle.
The merge tree is a tree that tracks these changes.
Its leaves are the sinks of the terrain, 
and its internal nodes are the negative saddles.
The edges of $\mergetree$, the merge tree, are in one-to-one correspondence with
the maximal depressions of $\terrain$, that is, we associate each edge $(u, v)$
with the maximal depression delimited by $u$ and containing $v$. 
See Figure \ref{fig:terrain} for an example. 
We assume that $\mergetree$ has an edge from the root of $\mergetree$ 
extending to $+\infty$, corresponding to the depression that extends to $\infty$. 
For simplicity, we assume that $\mergetree$ is binary,
that is, each negative saddle delimits exactly two depressions.
Non-simple saddles can be unfolded into a number of simple saddles
\cite{EHZ01}.

$\mergetree$ can be computed in $\BigO(n \log n)$ time~\cite{CSA03}, and it can be preprocessed in $O(n)$ 
additional time so that for a point $x \in \reals^2$, $\volume(\beta_x)$, the volume 
of the depression delimited by $x$ can be computed in 
$O(\log n)$ time~\cite{CSA03}. 
In the I/O model, $\mergetree$ can be constructed using $\BigO(\Sort(n))$ I/Os and as shown in~\cite{AR09},
the volume $\volume(\beta_x)$ and the smallest maximal depression containing $x$ can be computed for all vertices $x$ using $\BigO(\Sort(n))$ I/Os.


\subsection{Flooding Model}
We now describe the flooding model, which is the same as in \cite{lowe2019flood}, and define flow-rate functions.

\header{Flow graph and flow functions.}
We transform $\mesh$ into a directed acyclic graph $\flowgraph$, 
referred to as the \textit{flow graph}, by directing each edge $\{u,v\}$ of $\mesh$ 
from $u$ to $v$ if $h(u) > h(v)$, and from $v$ to $u$ otherwise, i.e., 
each edge is directed in the downward direction. 
For each (directed) edge $(u,v)$, we define the \emph{local flow} $\localflow(u,v,t)$ 
to be the portion of water arriving at $u$ that flows along the edge $(u,v)$ to $v$ at time $t$. 
By definition for any $u \in \vertices$, $\sum_{(u,v) \in \mesh} \localflow(u,v,t) = 1 $

The value of $\localflow(u,v,t)$ is, in general, based on relative heights of 
the downslope neighbors of $u$. 
If $u$ is not a negative saddle vertex, then $\localflow(u,v,t)$ remains the same for all $t$, 
so we will often drop $t$ and write $\localflow(u,v)$ to denote $\localflow(u,v,t)$. 
If $u$ is a negative saddle, then $\localflow(u,v)$ changes when one of its depressions fills up as no water flows 
for $u$ to that saddle; see below for further discussion. 

\header{Rain distribution.}
Let $\varregion(v,t): \vertices \times \reals \to \reals_{\geq 0}$ denote a \textit{rain distribution}, that is, for each vertex $v \in \vertices$, 
$\varregion(v,t)$ indicates the rate at which the rain is falling on $v$ at time $t$. 
We assume that for a each $v$, $\varregion(v,\cdot)$ is piecewise-constant function of time, with the function changing at discrete time values $\left\{ t_0=0,t_1,\dots,t_k \right\}$, and for all $v$ and $t \geq t_{k}$, $\varregion(v,t) = 0$. 
For a depression $\beta$, we define $\varregion(\beta,t) = \sum_{v \in \beta} \varregion(v,t)$. 
For $i \leq k$, let $|\varregion_i|$ denote the number of vertices for which $\varregion(v,t_i) \geq 0$, 
and let  $|\varregion| = \sum_{i=1}^k |\varregion_i|$ .
In practice $|\varregion_i| \ll n$. 


\header{Fill and spill rates.}
For a maximal depression $\beta$, we define the \textit{fill rate} 
$\fillrate_{\beta}: \reals_{\geq 0} \to \reals_{\geq 0}$ as the rate at which water is arriving in the depression $\beta$ 
as a function of time. 
That is, the rate at which rain is falling directly in $\beta$ 
plus the rate at which other depressions are spilling water into $\beta$. 
The fill rate $\fillrate_{\beta}$ is a monotonically nondecreasing piecewise-constant function. 
Similarly, we define the \textit{spill rate} $\spillrate_\beta: \reals_{\geq 0} \to [0,1]$ 
as the rate (as a function of time) at which water spills from $\beta$ through the saddle that delimits $\beta$.

\header{Flow rate.}
Next we define \textit{flow rates} $\flowrate_{e}$  and $\flowrate_{v}$ for edges $e$ and vertices $v$ of $\mesh$, 
which is the amount of water flowing through $e$ and $v$, respectively, at time $t$. 
For an edge $(u,v) \in \mesh$,  $\flowrate_{(u,v)}(t)$ is the fraction of water from $u$ that passes along $(u,v)$ as a function of time.
That is,
\begin{align}
    \label{eqn:eflowrate}
    \flowrate_{(u,v)}(t) &= \localflow(u,v,t)\flowrate_{v}(t).
\end{align}
The flow-rate $\flowrate_v$ of a non-saddle vertex $v$ is the sum of the flow-rates along incoming edges to $v$ plus the rain on the vertex $v$. That is, 
\begin{equation}
    \label{eqn:vflowrate} 
    \flowrate_v(t) =  \initrain(v,t)+\sum_{(u,v) \in \mesh} \flowrate_{(u,v)}(t).\\  
\end{equation}

Letting $\filltime_v$ be the time at which a vertex $v$ becomes flooded, $\flowrate_v(t)$  
and $\flowrate_{(u,v)}(t)$ for any $v \in \mesh$ are undefined for $t \geq \filltime_v$.
That is to say, when a vertex is flooded, the flow-rate function is undefined. 

Let $v$ be a negative saddle delimiting depressions $\alpha$ and $\beta$. 
Until one of $\alpha$ or $\beta$ is filled, $\flowrate_{v}$ is defined using equation~$\ref{eqn:vflowrate}$. 
Without loss of generality assume that depression $\alpha$ fills first, say at time $\filltime_\alpha$, 
and water starts spilling from $\alpha$ to $\beta$ through $v$. 
The spill rate $\spillrate_{\alpha}$ specifies the rate at which water spills from $\alpha$ to $\beta$. 
It is tempting to simply add $\spillrate_\alpha$ to equation~\ref{eqn:vflowrate}, 
but it double counts the amount of water that was flowing from $v$ to depression $\alpha$. 
For $t < \filltime_{\alpha}$, $\flowrate_{v}$ is defined as in (\ref{eqn:vflowrate}).
For $t \geq \filltime_\alpha$, $\flowrate_{v}$ is defined as follows, 
\begin{equation}
    \label{eqn:saddle-flow-rate}
    \flowrate_{v}(t) = \biggl( \varregion(v,t)+\sum_{(u,v) \in \mesh} \flowrate_{(u,v)}(t) \biggr)\sum_{w \in \beta} \localflow(v,w,0)+\spillrate_{\alpha}(t).
\end{equation}
Finally for $t \geq \filltime_{\alpha}$ and for any $w \in \beta$, 
\begin{equation}
    \label{eqn:saddle-edge} 
    \localflow(v,w,t) = \frac{\localflow(v,w,0)}{\sum_{z \in \beta} \localflow(v,z,0)}.
\end{equation}


\section{Terrain-flow Query}
\label{sec:mfd-flow}
In this section we describe an internal-memory algorithm that, given a terrain $\terrain$ and rain distribution $\varregion$, computes the flow-rate functions $\flowrate_{(u,v)}(t)$ for all edges $(u,v) \in \mesh$.

Each function is piecewise constant and represented as a sequence of pairs $(t_i,\Delta_i)$, where $\Delta_i$ represents the change in the function value at time $t_i$. 
We denote $|\flowrate_{(u,v)}|$ to be the combinatorial complexity of $\flowrate_{u,v}(t)$, i.e.,  the number of pieces defining $\flowrate_{(u,v)}$. 
Set $|\flowrate| = \sum_{(u,v) \in \edges} |\flowrate_{u,v}|$. 
If the rain distribution changes at $\BigO(1)$ times, 
then $|\flowrate| = \BigO(n \varheight)$, where $\varheight$ is the height of the merge tree. 

The preprocessing step builds, in $\BigO(n \log n)$ time, the merge tree $\mergetree$ of $\terrain$,
and labels each node in $\mergetree$ according to its in-order traversal as described in \cite{lowe2019flood}.
Furthermore, it computes $\volume(\beta_v)$ for each vertex $v \in \terrain$ and augments each
edge $(u,v) \in \terrain$ with the index of the smallest maximal depression
containing $v$.

We first present a simple algorithm and then describe how to adapt it to make it output-sensitive. 

\subsection{A simple algorithm}

We first give an overview of the algorithm. 
We sweep through the vertices of $\mergetree$ in descending height order, maintaining the following values for each height $\ell$:

\begin{enumerate}
    \item the set of depressions $\alpha_i$ in the sublevel set $h_{< \ell}$;

    \item for each $\alpha_i$, maintain the fill rate $\fillrate_{\alpha_i}(t)$, and

    \item for each $\alpha_i$ and the set of edges $(u,v) \in \mesh$ with $h(v) < \ell \leq h(u)$ and $v \in \alpha_i$,  denoted by $\edgecross(\alpha_i)$, maintain the flow-rate function $\flowrate_{(u,v)}(t)$.
\end{enumerate}

Note that $\fillrate_{\alpha_i}$ depends only on the sinks contained in $\alpha_i$, so we compute fill-rate functions of only maximal depressions. 
At each non-saddle vertex $v$ we compute $\flowrate_{v}$ and $\flowrate_{(v,w)}$, for each edge $(v,w) \in \mesh$,  
in a straightforward manner using (\ref{eqn:eflowrate}) and (\ref{eqn:vflowrate}). 
If $v$ is a negative saddle delimiting two depressions $\alpha$ and $\beta$, we must account for any rain spilling from one depression to the other. 
We do so by first partitioning the edges into two sets, $\edgecross(\alpha)$ and  $\edgecross(\beta)$.
Then given the flow-rates on the edges in each set, we compute the fill-rates $\fillrate_{\alpha}$ and $\fillrate_{\beta}$. 
Given these fill-rates, we then determine which depression (if any) fills first. 
If at least one depression becomes full, assume that $\beta$ fills first without loss of generality. 
Then we can compute the spill-rate function $\spillrate_{\beta}(t)$. 
Finally, we compute $\flowrate_{(v,w)}$ for each edge $(v,w) \in \mesh$.  

We will now describe the algorithm in detail. We begin by augmenting $\mergetree$ with additional information which will be used when computing the fill-rates. 

Given a rain distribution $\varregion$, before performing the downward sweep, 
we first compute $\varregion(\beta,\cdot)$ for each maximal depression $\beta$ as follows:
first assign $\varregion(v,\cdot)$ for each vertex with nonzero rainfall to the smallest maximal depression containing $v$, and then perform an upward sweep on the merge-tree $\mergetree$, maintaining the sum of rainfall functions at each saddle vertex. 

As we sweep top-down, for each depression $\alpha_i$ we store the flow-rates of the edges in $\edgecross(\alpha_i)$ in a data structure that supports the following three operations:
(i) add $\flowrate_{(u,v)}(t)$ to the data structure; (ii) remove $\flowrate_{(u,v)}(t)$ from the data structure; (iii) and given a saddle vertex delimiting two depressions $\beta,\gamma \in \alpha_i$, 
partition the flow-rates into two sets $\edgecross(\beta)$ and $\edgecross(\gamma)$ and return the sum of flow-rates in each set.

Noting there are $\BigO(\varheight+k)$ values of $t$ at which the flow-rate functions can change
(corresponding to the times at which $\varregion$ changes and spill-times of tributaries of the depression in question) we build the data structure as follows. 
For each time $t_i$ at which the functions can change, 
maintain the values $\Delta_i$ for the flow-rate of each edge $(u,v)$ sorted according to the label $\ell(v)$. 
Additionally maintain the prefix sum of the values for each $t_i$. 
There will be $\BigO(|\flowrate_{(u,v)}|)$ values to consider at each edge, so we can insert and remove flow-rates of edges in $\BigO( |\flowrate_{(u,v)}| \log n)$-time. 

To partition the edges into $\edgecross(\beta)$ and $\edgecross(\gamma)$, 
recall each vertex is labeled according to an in-order traversal of $\mergetree$. 
Given two depressions $\beta$ and $\gamma$ delimited by a saddle vertex $v$ with label $\ell(v)$, 
all vertices contained in $\beta$ will have a label less than $\ell(v)$, 
and all vertices contained in $\gamma$ will have a label greater than $\ell(v)$. 
Thus, to partition the edges into the depressions they cross into we simply take the prefix of all edges 
with label less than $\ell(v)$ to be one set, and the remaining suffix to be the other set. 
Then given the partition, we can use the prefix sums to determine the sum of flow-rate functions into $\beta$ and $\gamma$ in $\BigO(\varheight+k)$ time.

With this data structure at hand, we now describe how to process each vertex $v$ as we encounter it in our sweep.

\header{Non-negative-saddle vertex.}
If $v$ is a non-negative-saddle vertex computing $\flowrate_{v}$ is easy using (\ref{eqn:vflowrate}).
Note that $\flowrate_{(u,v)}$ for all incoming edges to $v$ has been computed. 
Additionally, using $\fillrate_{\alpha_i}(t)$, for the maximal depression $\alpha_i$ containing $v$, along with $\volume(\beta_v)$, we can determine if or when $v$ becomes flooded.

\header{Negative saddles.} 
If $v$ is a negative saddle delimiting two depressions $\alpha$ and $\beta$, 
we first determine whether either of $\alpha$ or $\beta$ becomes full and, if so, which fills first. 
The fill rate of a depression $\alpha$ is the sum of the rain falling directly on vertices in $\alpha$ plus the flow-rates across all edges crossing into $\alpha$, that is
\begin{equation}
    \label{eqn:saddlefill}
    \fillrate_{\alpha}(t) = \sum_{u \in \alpha} \varregion(u,t) +\sum_{(v,w) \in \edgecross(\alpha)} \flowrate_{(v,w)}(t).
\end{equation}

The first sum has already been computed in the upward sweep of the merge tree. 
However, to compute $\flowrate_{(v,w)}(t)$ for all values of $t$ we would need to know which depression fills first and when. 
So instead, we define the \textit{pseudo-flow-rate function}, $\flowrate'_{(v,w)}$,  in the same manner that flow-rate for non-saddle vertices is computed as in (\ref{eqn:vflowrate}). 
Using this, we compute a modified \textit{pseudo-fill-rate} function  
\begin{equation}
    \label{eqn:pseudosaddlefill}
    \fillrate'_{\alpha}(t) = \sum_{u \in \alpha} \varregion(u,t) +\sum_{(v,w) \in \edgecross(\alpha)} \flowrate'_{(v,w)}(t).
\end{equation}

Before $\alpha$ or $\beta$ become full (i.e. $t < \min(\filltime_\alpha, \filltime_\beta)$) we have that $\flowrate'_{(v,w)}(t) = \flowrate_{(v,w)}(t)$, which in turn implies that $\fillrate'_{\alpha}(t) = \fillrate_{\alpha}(t)$ (resp. $\fillrate'_{\beta}(t) = \fillrate_{\beta}(t)$) before $\alpha$ or $\beta$ becomes full. 
Given this, compute $\flowrate'_{(v,w)}(t)$ for all edges from the saddle $v$ and add these flow-rates to the data structure. 
We then use the data structure to partition the edges and compute the pseudo-fill-rates of the two depressions,   
from which we can determine which (if any) depression fills first. 

If neither $\alpha$ nor $\beta$ becomes full, we are done. 
If not, assume without loss of generality that $\alpha$ becomes full first and spills into $\beta$. 
Then given $\fillrate_{\alpha}(t)$ we compute $\spillrate_{\alpha}(t)$ and in turn compute $\flowrate_{v}(t)$ for $\filltime_\alpha \leq t$ using (\ref{eqn:saddle-flow-rate}) along with $\fillrate_{\beta}(t)$ using (\ref{eqn:saddlefill}). 

Computing $\varregion(\beta,\cdot)$ for all maximal depressions can be done in 
$\BigO(|\varregion|+\varsinks k)$-time where $\varsinks$ is the number of sinks in the terrain 
and $k$ is the number of times the rain distribution changes.
To compute the flow-rate functions of edges from all non-negative saddle vertices, along with the pseudo-fill-rate functions from negative saddles, we spend $\BigO(|\flowrate| \log n)$-time computing the flow-rate functions and adding them to the data structure. 
At a negative saddle delimiting two depressions $\alpha,\beta$, if we partition the edges by walking from the front and back of the sorted list of edges we spend $\BigO(\min(|\alpha|,|\beta|)$-time, where $|\alpha|$ is the number of vertices contained in  $\alpha$. 
A simple recurrence shows that the total time spent partitioning edges at all negative saddles is $\BigO(n \log n)$. 
Finally computing the fill and spill rates at a negative saddle takes time proportional to their complexity, which is $\BigO(k+\varheight)$. Noting that $|\flowrate| = \BigO(n(\varheight+k))$, we get the following.

 \begin{theorem}
     \label{thm:simple-terrain-flow}
     Given a triangulation $\mesh$ of $\reals^2$ with $n$ vertices and a height function $h: \mesh \to \reals$ that is linear on each face of $\mesh$, a data structure of size $\BigO(n)$ can be constructed in $\BigO(n \log n)$ time so that for a (time varying) rain distribution $\varregion$, 
     a terrain-flow query can be answered in $\BigO(\varsinks k+n(\varheight+k)\log n)$ time, where $\varsinks$ is the number of sinks in $\mesh$, $k$ is the number of times at which the rain distribution changes.
 \end{theorem}

     In particular, if $k = \BigO(n)$, we have that the terrain-flow query can be answered in $\BigO(n^2 \log n)$ time. 

\subsection{An Output-Sensitive Algorithm}
We modify the above algorithm so that its running time becomes $\BigO(\varsinks k+|\flowrate| \log n)$ 
where $|\flowrate|$ is the total complexity of all non-zero flow-rate functions. 
If water flows across a small portion of $\terrain$ then $|\flowrate| \ll n$ and the new algorithm will be faster. 
The key idea is we do not have to process vertices with zero flow-rate. 

The algorithm begins in the same manner as the previous algorithm, computing $\varregion(\beta,\cdot)$ for each maximal depression $\beta$. 

However rather than sweeping over all vertices of $\mergetree$, 
we instead maintain a priority queue of vertices with nonzero flow-rate with their heights as their priority, i.e., the larger the height, the higher the priority. 
Every vertex $v$ with nonzero flow-rate will either have rain falling directly on it, 
or have a path in $\flowgraph$ from a source vertex (i.e. either a vertex with rain falling directly on it, or a saddle from which water is spilling.) 
We initialize the priority queue with all vertices on which rain falls directly along with all negative saddles. 

While the priority queue is not empty, we pop the highest priority vertex $v$ from the queue and compute the flow-rate function $\flowrate_v(t)$, as we have described in the previous algorithm. 
For each vertex $w$ adjacent to $v$ in $\flowgraph$ with $\flowrate_{(v,w)} > 0$, 
add $w$ to the priority queue. 
For each edge, in addition to each value $\Delta_i$ representing the change in the flow-rate function at time $t_i$, 
 we also store $V_i$, the integral of the flow-rate function from $t=0$ to $t=t_i$. 
 These are also stored in sorted order, and we maintain the prefix sum of these values for each edge. 

At negative saddles, to determine which depression fills first, when we partition the edges we also compute the sum of volumes at the last time step $t_{j}$ for the depressions $\alpha$ and $\beta$ delimited by the saddle vertex.
 If both have filled by this time, consider the sum at $t_{j-1}$ and so on until we find the latest time at which only one depression is filled. 
 This will be the depression which fills first, and ensures the number of time steps considered will be $\BigO(|\spillrate_{\beta}|)$. 
 Then we compute the fill and spill-rates as we did in the prior algorithm. 

 Since we compute the flow and spill-rates as before, it suffices to prove that we consider each negative saddle with positive spill-rate and each vertex with positive flow-rate. 
 We begin with all negative saddles in the priority queue, so the former holds trivially. 
 For the latter, given a vertex $v$ with positive flow-rate, either rain falls directly on it, 
 in which case we added $v$ when initializing the priority queue, 
 or water flows to it from some higher vertex $u$. 
 If $u$ is added to the priority queue, when it is processed $v$ will in turn be added and later processed. 
 Since all non-source vertices with positive flow-rate have a path from source vertex, 
 and all source vertices begin in the priority queue, we have that we will reach and process $v$. 

The time spent processing non-saddle vertices $v$ and edges $(v,w)$ will be $\BigO(|\flowrate_{v}(t)|)$ and 
$\BigO(|\flowrate_{(v,w)}(t)|)$ respectively. 
So the total time processing these will be $\BigO(|\flowrate|)$. 
At negative saddle vertices when we partition the edges, if we perform the search by walking from the front and back of the list we will spend a total of $\BigO(|\flowrate| \log |\flowrate|)$ time partitioning edges at all negative saddles. 
Finally, at each negative saddle $v$ which is a source we spend $\BigO(|\spillrate_{\beta}|)$-time computing which depression fills first. 
Since $|\spillrate_{\beta}| \leq |\flowrate_{v}|$, the total time processing saddle vertices is $\BigO(\rho k+|\flowrate_{v}(t)| \log n)$. 
 \begin{theorem}
     \label{thm:terrain-flow}
     Given a triangulation $\mesh$ of $\reals^2$ with $n$ vertices and a height function $h: \mesh \to \reals$ that is linear on each face of $\mesh$, a data structure of size $\BigO(n)$ can be constructed in $\BigO(n \log n)$ time so that for a (time varying) rain distribution $\varregion$, 
     a terrain-flow query can be answered in $\BigO(\varsinks k+|\flowrate| \log n)$ time, where $\varsinks$ is the number of sinks in $\mesh$, $k$ is the number of times at which the rain distribution changes, and $|\flowrate|$ is the total complexity of all non-zero flow-rate functions. 
 \end{theorem}

\section{I/O-Efficient Algorithms}
\label{sec:io}
In this section we describe two I/O-efficient algorithms that given a terrain
$\terrain$ and a rain distribution $\varregion$ determine the flow-rate
$\flowrate_{(u,v)}(t)$ for all edges $(u,v) \in \terrain$.
The algorithms use the same framework as described in \rsec{mfd-flow}.
However, we cannot explicitly maintain the list of edges crossed by the sweep
line in memory due to the bounded internal memory size.
We instead store the merge tree in memory and rely on the
\textit{time-forward processing} technique, where the flow-rates computed at
vertices are forwarded to downslope neighbors using an I/O-efficient priority
queue~\cite{CGGTVV95}.
The I/O-efficient priority queue supports $n$ insertions and deletions in
$\BigO(\Sort(n))$ I/Os and $\BigO(n \log n)$ internal computation
time~\cite{sanders2000queue}.


In the preprocessing step of both algorithms, we compute the merge tree
$\mergetree$ of $\terrain$ and label each node in $\mergetree$ according to
its in-order traversal as described in~\cite{lowe2019flood}.
Furthermore, we compute $\volume(\beta_v)$ for each vertex $v \in \terrain$ and augment each
edge $(u,v) \in \terrain$ with the index of the smallest maximal depression containing $v$.
This can be computed in $\BigO(\Sort(n))$ I/Os using the algorithm described by
Arge et al.~\cite{AR09}. Since both algorithms assume $\varsinks = \BigO(m)$, we assume the $\mergetree$ is stored in internal memory.

\subsection{Extending the Internal Memory Algorithm}
We now describe our first I/O-efficient algorithm.
In order to extend the internal terrain-flow query algorithm, we introduce the
following notation: let $\edgeuniquecross(\alpha) \subseteq \edgecross(\alpha)$
be the set of edges such that, for each edge $(u,v) \in
\edgeuniquecross(\alpha)$, $\alpha$ is the smallest maximal depression
containing $v$.

As in the previous section, we proceed by performing an
upward sweep on the terrain followed by a downward sweep.

\header{Upward sweep.}
For the upward sweep of our algorithm, we compute
$\varregion(\alpha,\cdot)$ for each maximal depression $\alpha$.
Given a rain distribution $\varregion$, we start by computing the sum of rain
falling directly in each maximal depression $\alpha$.
We do this by assigning $\varregion(v, \cdot)$ for each vertex with nonzero
rainfall to the smallest maximal depression containing $v$.
This step can be implemented I/O-efficiently using a sort and a scan
of the terrain and $\varregion$;
for each maximal depression $\alpha$ store $\varregion(\alpha, \cdot)$ in memory.
We then perform the upward sweep described in \rsec{mfd-flow} in memory,
maintaining the sum of rainfall functions at each saddle vertex.
This upward sweep can be trivially implemented in $\BigO(\Sort(|\varregion|
+ n))$ I/Os and $\BigO(|\varregion| \log |\varregion| + n \log n + \varsinks k)$ internal computation time.

\header{Downward sweep.}
We sweep through vertices of $\terrain$ in descending height order,
maintaining the following values in memory for each height $\ell$:
\begin{enumerate}
  \item for each depression $\alpha_i$ in the sublevel set $h_{<\ell}$, maintain the fill-rate $\fillrate_{\alpha_i}(t)$, and
  \item for each $\alpha_i$ maintain the \textit{depression flow-rate sum}\\ $\sum_{(u,v) \in
\edgeuniquecross(\alpha)}\flowrate_{(u,v)}(t)$
\end{enumerate}
We cannot explicitly store the flow-rates for the set of edges crossing the
sweep line in memory.
Instead, we store each flow-rate $\flowrate_{u,v}(t)$ in an I/O-efficient
priority queue $Q$ keyed on the height of vertex $v$.
Furthermore, we initialize $Q$ by inserting the
functions $\varregion(v,t)$ for each vertex $v$ with $\varregion(v) > 0$.
Whenever we process a vertex $v$, for all $(v,u) \in \terrain$ we propagate the
flow-rate $\flowrate_{(v,u)}$ forward to $u$ using 
$Q$ such that the function can be removed from $Q$ whenever $u$ is processed.
The values maintained for each depression $a_i$ in sublevel set $h_{<\ell}$ can
be maintained in memory since we assume $\varsinks (\varheight + k) = \BigO(\varmem)$.

We now describe how each step of the sweep is performed.
Let $v$ be the current vertex.
First we remove the function $\varregion(v, t)$ from $Q$.
Furthermore, for each edge $(u,v) \in \terrain$, we remove the flow-rate
$\flowrate_{(u,v)}(t)$ from $Q$.
Since the algorithms visits vertices in descending order of height, the
functions must have been inserted into $Q$ at an earlier point and will be at the front
of $Q$.
Since each edge $(u,v)$ no longer crosses the sweep line, we update the
depression flow-rate sums by subtracting $\flowrate_{(u,v)}(t)$ from the
flow-rate sum of the smallest maximal depression containing $v$.

Let $\alpha_i$ be the smallest maximal depression containing $v$.
If $v$ is a non-negative-saddle vertex, we compute $\flowrate_v$ and
determine if or when $v$ becomes flooded, using $\fillrate_{\alpha_i}(t)$ and
$\volume(\beta_v)$ as described in \rsec{mfd-flow}.

If $v$ is a negative saddle delimiting two depressions $\alpha$ and $\beta$, we
wish to compute whether $\alpha$ or $\beta$ becomes full and, if so, which
fills first.
In order to compute the flood times of the two depressions, we recall that the
fill-rate of a depression $\alpha$ is equal to the sum of flow-rates across all
the edges $(u,w) \in \edgecross(\alpha)$ plus the amount of rain falling in
$\alpha$.
When we are processing the vertex $v$ we have that all edges crossing into
$\alpha$ and $\beta$ respectively are also crossing the sweep line.
We therefore compute the fill-rate of $\alpha$ ($\beta$ resp.) by summing the
depression flow-rate sums for all maximal depressions $\alpha_i \subseteq
\alpha$ ($\beta_i \subseteq \beta$ resp.):
\begin{align}
  \fillrate_\alpha(t) 
                     &= \varregion(\alpha, t) + \sum_{\alpha_i \subseteq \alpha} \sum_{(u,v) \in \edgeuniquecross(\alpha_i)}\flowrate_{(u,v)}(t).
\end{align}
The flood times and the flow-rate $\flowrate_{v}(t)$ is then computed from
$\fillrate_\alpha(t)$ and $\fillrate_{\beta}(t)$ as described in \rsec{mfd-flow}.

Finally, we can propagate the flow-rates on the outgoing edges by pushing
$\flowrate_{(v,w)}(t) = w_{(v,w)} \cdot \flowrate_v(t)$ to $Q$
for all vertices $w$ where $w_{(v,w)} > 0$.
Furthermore, we update the depression flow-rate sums with the flow-rates of the
outgoing edges.
Note that each flow-rate is added to only one flow-rate sum.

The flow-rate $\flowrate_{(v,w)}(t)$ of each edge $(v,w)$ is forwarded
only once using the priority queue.
Thus, we spend a total of $\BigO(\Sort(|\flowrate|))$ I/Os and
$\BigO(|\flowrate| \log |\flowrate|)$ internal computation time processing all
vertices in the sweep excluding computation of the fill-rates at negative
saddles.
In order to efficiently compute the fill-rates in internal memory, we store
the depression flow-rate sums in a sorted list ordered by the in-order
traversal index of the depressions.
We then partition the sums as in \cite{lowe2019flood} and compute the
fill-rates using $\BigO(|\flowrate| \log |\flowrate|)$ internal computation
time in total.
Since the initial sorting of vertices and edges can be performed in
$\BigO(\Sort(n))$ I/Os and $\BigO(n \log n)$ internal computation time, the
algorithm uses a total of $\BigO(\Sort(|\flowrate|))$ I/Os and
$\BigO(|\flowrate| \log |\flowrate|)$ internal computation time.

\begin{theorem}
  \label{thm:io-terrain-flow}
  Given a triangulation of $\mesh$ with $n$ vertices, a height function $h:
  \mesh \to \reals$ which is linear on each face of $\mesh$ and a rain
  distribution $\varregion$, a
  terrain-flow query can be answered in $\BigO(\Sort(|\flowrate|))$ I/Os and
  $\BigO(|\flowrate| \log |\flowrate|)$ internal computation time assuming $\varsinks (\varheight + k) = \BigO(\varmem)$, where
  $|\flowrate|$ is the total complexity of all flow-rate functions which we
  return, $\varheight$ is the height of the merge tree, $k$ is the number of times
  at which the rain distribution changes, $\varsinks$ is the number of sinks in
  $\mesh$, and $\varmem$ is the size of internal memory.
\end{theorem}

\subsection{Assuming Smaller Internal Memory}
We now extend the algorithm to relax the assumption on the size of the
internal memory from $\varsinks (\varheight + k) = \BigO(\varmem)$ to
$\varsinks = \BigO(\varmem)$, at the cost of a greater number of I/Os.

We use the same framework as described previously, however, we avoid storing
the depression flow-rate sums and fill-rates in memory for each depression in
the sublevel set $h_{< \ell}$;
We instead the priority queue $Q$ to forward fill-rates as well as
the edge flow-rates used to compute fill-rates at negative saddle vertices.

\header{Forwarding fill-rates.}
Let $v$ be non-negative-saddle vertex and let $\alpha_i$ be the smallest maximal depression
containing $v$.
Let $u$ be the vertex visited after $v$ in the downward sweep, where the
smallest maximal depression containing $u$ is also $\alpha_i$.
When performing the sweep, we forward $\fillrate_{\alpha_i}(t)$ from $v$ to $u$
using $Q$.
We note that we can augment $v$ with the height of $u$ using $\Sort(n)$ I/Os in
preprocessing, and thus we can forward $\fillrate_{\alpha_i}(t)$ to $u$ during the
sweep.
Furthermore, for each negative saddle vertex $v$ delimiting depressions
$\alpha$ and $\beta$, we forward $\fillrate_\alpha(t)$ and
$\fillrate_\beta(t)$ to the first vertices visited in $\alpha$ and $\beta$,
respectively.

\header{Computing fill-rates at negative saddles.}
Let $v$ be a negative saddle vertex delimiting two depressions $\alpha$ and
$\beta$.
During execution of the sweep, we wish to compute the fill-rates of depressions
$\alpha$ and $\beta$.
We recall that the fill-rate of $\alpha$ can be computed as follows:
\begin{align}
  \fillrate_\alpha(t) &= \varregion(v, t) + \sum_{(u,v) \in E(\alpha)} \flowrate_{(u,v)}(t).
\end{align}
We note that the flow-rates required to compute this sum and $\varregion(v, t)$
can be propagated using $Q$.
However, that would in the worst-case lead to forwarding $\BigO(n \varheight)$
functions in total.
We recall that $\fillrate_{\beta_v}(t)$ is forwarded to $v$ using $Q$.
Furthermore, since $\fillrate_{\beta_v}(t) = \fillrate_{\alpha}(t) +
\fillrate_{\beta}(t)$, it suffices to compute either $\fillrate_\alpha(t)$ or
$\fillrate_{\beta}(t)$, whichever requires the fewest flow-rates to be
forwarded.
The number of flow-rate functions that need to be forwarded in order to
compute $\fillrate_{\alpha}(t)$ and $\fillrate_{\beta}(t)$, respectively, can
be precomputed by counting the number of edges crossing the boundaries of
$\alpha$ and $\beta$.
This precomputation step can trivially be implemented by performing a scan of
the vertices using $\BigO(\Scan(n))$ I/Os and $\BigO(\varsinks)$ memory.
We therefore preprocess for which depressions we compute fill-rates and forward
only the flow-rates required for computing those.

We now bound the number of edges forwarded using a similar recurrence as
\cite{lowe2019flood};
Let $|\alpha|$ denote the total number of edges with at least one vertex
contained in the depression $\alpha$ and note that $|\alpha| \geq
|\edgecross(\alpha)|$.
Letting $T(\beta_v)$ be the total number of flow-rates summed to compute the
fill-rates for all saddles contained in $\beta_v$, we have that
\begin{align}
  T(\beta_v) = \BigO \big(\min(|\alpha|, |\beta|)\big) + T(\alpha) + T(\beta).
\end{align}
Noting that the set of edges crossing into the two maximal depressions $\alpha$
and $\beta$ are disjoint, it follows that $|a| + |\beta| \leq |\beta_v|$.
Using this, the recurrence solves to $T(\beta_v) = \BigO(|\beta_v| \log
|\beta_v|)$, and we thus need to forward only a total of $\BigO(n \log n)$
flow-rate functions.
Since the complexity of each flow-rate function is bounded by $\BigO(\varheight
+ k)$, we spend a total of $\BigO(\Sort((\varheight + k) n \log n))$ I/Os and
$\BigO((\varheight + k) n \log^2 (k n))$ internal computation forwarding edges
and computing fill-rates.

\begin{theorem}
  \label{thm:io-terrain-flow-2}
  Given a triangulation of $\mesh$ with $n$ vertices, a height function $h:
  \mesh \to \reals$ which is linear on each face of $\mesh$ and a rain
  distribution $\varregion$, a terrain-flow
  query can be answered in $\BigO(\Sort((\varheight + k) n \log n))$ I/Os and
  $\BigO((\varheight + k) n \log^2 (k n))$ internal computation time assuming
  $\varsinks = \BigO(\varmem)$, where $\varheight$ is the height of the merge tree,
  $k$ is the number of times at which the rain distribution changes, $\varsinks$ is the
  number of sinks in $\mesh$, and $\varmem$ is the size of internal memory.
\end{theorem}


\section{Edge-Flow Query}
\label{sec:pointflow}
The terrain-flow query can naturally be used to answer edge-flow queries by returning $\flowrate_{e}(t)$ for a query edge $e = (q,r) \in \mesh$. 
While in practice the query time can be improved, the worst case running time under the MFD model remains the same as the terrain-flow query. 
Under the SFD model, we can improve this running time significantly, building on the fast algorithm for the flood-time query under SFD given by Rav \etal~\cite{RLA17}, along with a linear-size data structure supporting constant time reachability queries in planar directed graphs given by Holm \etal~\cite{holm2015planar}.

The key idea of the algorithm is that under the SFD model when water falls on a vertex or spills from a negative saddle the water flows along a single path to some sink in the terrain. 
Thus if we can find which vertices and negative saddles from which water follows a path containing $(u,v)$, 
$\flowrate_{(u,v)}$ will be the sum of water falling directly on or spilling from these sources. 
Before describing the algorithm, we begin with some definitions. 

Let $u$ be a negative saddle, let $(u,v_1)$ and $(u,v_2)$ be two down edges in $\mergetree$ from $u$, 
and let $(w,u)$ be the up edge from $u$. 
We call the depression associated with $(u,v_2)$ (resp. with $(w,u)$) as the \textit{sibling} (resp. \textit{parent}) (depression) of that associated with $(u,v_1)$. 
Any given point $q \in \mesh$ is contained in a sequence of maximal depressions
$\alpha_1 \supset \dots \supset \alpha_k \ni q$. 
Each $\alpha_i$ is delimited by a saddle $v_i$
and has a corresponding sibling depression $\beta_i$.
Note that these saddles form a path in $\mergetree$ from $q$ to the root.
We refer to the maximal depressions $\beta_1, \dots, \beta_{k-1}$
as the \emph{tributaries of $q$} and denote them by $\tribs_q$. 

For any point $q \in \mesh$ we define the \textit{tributary tree} $\tribdag_q$ as follows. $\tribdag_q$ is a directed graph with nodes corresponding to the tributaries of $q$ plus $\beta_q$. There is an edge $(\beta_i,\beta_j)$ in $\tribdag_q$ if water spills from the saddle $v_i$ to a sink in $\beta_j$ when $\beta_i$ becomes full. 
Water spills to exactly one sink under the SFD model, so $\tribdag_q$ will be a tree rooted at $\beta_q$.  

We present an $\BigO(n \log n)$-time algorithm for preprocessing $\terrain$ into a linear-size data structure that can answer an edge-flow query for a given  rain distribution $\varregion(t)$ and query edge $(q,r)$. The query takes $\BigO(|\varregion|+\tribs_{q} k \log n)$, where $\tribs_{q}$ is the number of $q$-tributaries $\beta$ with $\varregion(\beta, t) > 0$. 

Given a query edge $(q,r) \in \mesh$ and a rain distribution $\varregion(t)$, we begin by assuming that water flows from $q$ to $r$. 
Since water only flows from each vertex to one neighbor in the SFD model, if this were not the case then we would immediately have that $\flowrate_{(q,r)} = 0$. 
Hence, $\flowrate_{(q,r)} = \flowrate_{q}$. 
For simplicity, we will instead compute the equivalent point-flow query. 

\begin{figure}[t]%
  \centering
  \includegraphics[]{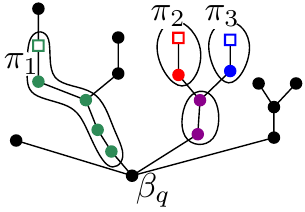}%
  \caption{%
    \normalfont
    The tributary tree $\tribdag_q$ rooted at $\beta_q$ with each vertex denoting a tributary of $q$, rain initially falls in the tributaries marked as squares. 
  }
  \label{fig:pathCompress}%
\end{figure}

The algorithm begins by building $\tribdag_q$. 
See Figure \ref{fig:pathCompress} for an example. 
As we have noted, water will reach $q$ in one of two cases: rain falls on a vertex $v$ of the terrain and follows a path which crosses $q$, 
or water spills from a tributary of $q$ and reaches $q$. 

Consider first the case when rain falls only on a single point $p$ contained in some tributary $\beta_{i_1}$. 
Take the path $\pi_1$ in $\tribdag_q$ from $\beta_{i_1}$ to the root $\beta_q$, $(\beta_{i_1}, \beta_{i_2}, \cdots, \beta_{i_k}, \beta_q).$
For each depression $\beta_{i_j}$ let $V_i$ denote the depression volume of $\beta_{i_j}$ and let $\tau_j$ be the fill-time of $\beta_{i_j}$. 
The fill-time $\tau_k$, when $\beta{i_k}$ begins spilling into $\beta_q$, will be when the volume of rain falling on $p$ equals $\sum V_i$. Moreover we have
$$
\fillrate_{\beta_q}(t) = \spillrate_{\beta_{i_k}}(t) = 
\begin{cases}
    0 &t < \tau_k\\
    \fillrate_{\beta_{i_1}}(t) &t \geq \tau_k.\\
\end{cases}
$$

That is to say, instead of computing the fill and spill rates for each tributary along the path, we can merge all the tributaries in this path and treat it as if it were a single depression.   
Then to answer the query, check whether there is a path from the saddle delimiting $\beta_{i_k}$ to the query vertex $q$. 
If there is, we have $\flowrate_{q}(t) = \spillrate_{\beta_{i_k}}(t)$, otherwise $\flowrate_{q}(t) = 0$.

Now consider the general case where rain falls on many vertices. 
We will compute $\flowrate_{q}$ as a sum of the rainfall functions on vertices that directly reach $q$, along with the spill-rates of parents of $\beta_q$ in $\tribdag_q$ that reach $q$. 
We begin by computing the initial fill-rate of each tributary in which rain falls directly. 
For each vertex $v$ from which rain flows to a sink contained in $\beta_q$, 
check whether the path the water takes crosses $q$. If so add their rainfall to the sum.
If rain falls in multiple tributaries that have disjoint paths in $\tribdag_q$ to $\beta_q$ (excluding the root $\beta_q$), we can simply perform the single-point rain algorithm multiple times for each path and add each spill-rates from depressions which reach $q$ to the sum. 
However it might be the case that two such paths intersect at a tributary $\gamma$ before reaching $\beta_u$ (e.g. $\pi_2$ and $\pi_3$ in Figure \ref{fig:pathCompress}.) 
Here, we can compute the spill-rates of the two parent tributaries of $\gamma$ as we did in the single-point rain algorithm, and then recurse, treating $\gamma$ as a single-point source of rain with fill-rate equal to the sum of spill-rates of its parent tributaries. 
Cases where more than two paths merge at a single vertex can be handled in a similar manner. 

Now it remains to show how we can perform this algorithm efficiently. 
There are two main operations needed. 
First, we must compute the fill and spill-rates of the tributaries of $q$. 
Then we must determine from which saddles and vertices water will reach $q$. 

To perform the first operation efficiently, build the linear-size fast data structure for flood-time queries as described in \cite{RLA17}. 
With this data structure we can compute the spill-rates of parent tributaries of $\beta_q$ in $\BigO(|\varregion|+\tribs_{q}k \log n)$ time, where $\tribs_q$ as defined above, is the numbe of $q$-tributaries in which rain is falling directly. 

To perform the second operation efficiently, we consider the subgraph of the flow graph $\flowgraph$ taking only edges for which $\localflow(u,v,0) > 0$, i.e., when $v$ is the lowest neighbor of $u$. 
Then on this directed planar graph, build the linear-size data structure reachability queries as described in \cite{holm2015planar}. 
This data structure supports constant-time reachability queries. 
For a non-saddle vertex $v$, water falling at $v$ reaches the query vertex $q$ if $q$ is reachable from $v$.
For water spilling from a negative-saddle $v$ towards a node $u$ we instead check whether $q$ is reachable from this vertex $u$. 
Omitting all the details, we obtain the following:

\begin{theorem}
    \label{thm:sfd}
    Given a triangulation $\mesh$ with $n$ vertices and a height function $h: \mesh \to \reals$ which is linear on each face of $\mesh$, a data structure of size $\BigO(n)$ can be constructed in $\BigO(n \log n)$ time so that for a (time varying) rain distribution $\varregion(t)$ and an edge $(q,r)$ an edge-flow query can be answered in $\BigO(|\varregion|+\tribs_{q} k \log n)$,  where $|\varregion|$ is the complexity of the rain distribution, $\tribs_{q}$ is the number of tributaries in which rain is falling directly, and $k$ is the number of times at which the rain distribution changes.
\end{theorem}


\section{Extracting 2D Flow Networks}
\label{sec:2dflow}
So far we have assumed that water flows along the edges of $\terrain$, and computed the flow rate of water along these edges. But in reality the water is not constrained to edges and flows along 2D channels forming a 2D network of rivers. 
In this section we first describe a model for determining the 2D channels given a 1D flow network. 
We then present an efficient algorithm for computing these channels for a given path along the edges of $\mesh$. 

\subsection{Model for 2D Channels}
We assume that we are given a path $\paths$ along the edges of $\mesh$. For each edge $e$ of $\paths$, let $\flowrate_{e} \in \reals_{\geq 0}$ be the flow value along $e$. 
Unlike previous sections, we assume that $\flowrate_{e}$ does not vary with time. 
But $\flowrate_{e}$ may vary with edges.  
The goal is to compute a 2D channel $\channel:= \channel(\paths, \flowrate)$ along which water flows.
\footnote{One method of generating $\paths$ is computing the flow rate along all edges of $\mesh$ as in Sections \ref{sec:mfd-flow} and \ref{sec:io}, fixing a time $t=c$, choosing a threshold $\psi$, extracting the edges $e$ with $\flowrate_{e}(c) \geq \psi$, post-processing these edges to construct a 1D flow network with some desired properties,  
and finally decomposing this flow network into a set of paths.  These steps are beyond the scope of this paper and an interesting direction for future research.}
The channel $\channel $ is defined by its left and right banks and the height of water at every point on $\paths$. 
More precisely, $\paths$ is parameterized as $\paths: I \to \reals^2$ where $I = [x_0,x_1]$ is an interval. 
For every $x \in I$, we define $\lb(x), \rb(x) \in \reals^2$ as the left and right bank, respectively, of $\channel$ at $x$, 
and $\Delta(x) = h(\lb(x)) = h(\rb(x)).$
The locus of points $\lb(x)$ (resp. $\rb(x)$), for $x \in I$, traces a curve $\lb$ (resp. $\rb$), which is the \textit{left} (resp. \textit{right}) \textit{bank} of $\channel$. 

\begin{figure}[t]%
  \centering
  \includegraphics[]{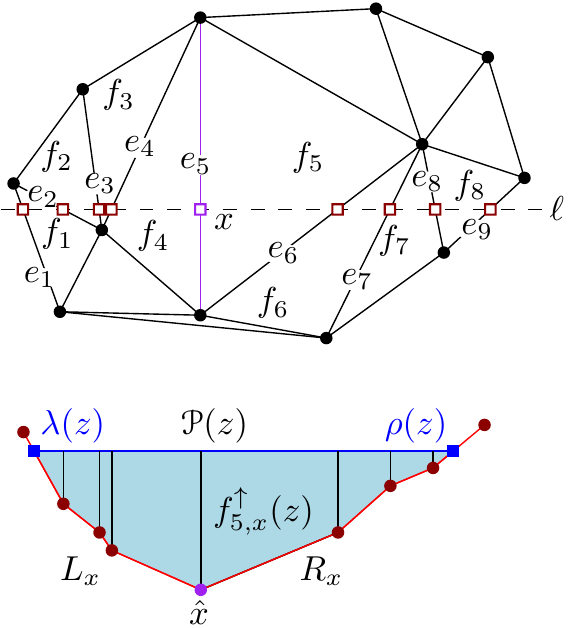}%
  \caption{%
    \normalfont
    Top: $\terrain$ with the edge $e_5$ marked in purple and line $\ell$. The intersection of $\ell$ with edges of $\terrain$ marked with boxes. 
    Bottom: $\terrain_{x}$, with water at height $z$. The polygon $\profile$ is marked in blue, and the wetted perimeter marked in red. The vertices correspond to the intersection point above in the top figure.
  }
  \label{fig:channel}%
\end{figure}

\footnote{Here we assume that $\lb$ and $\rb$ are simple curves; if either of them is self-intersecting, then $\channel$ has to be defined more carefully.} 
We overlay $\mesh$ with $\lb$ and $\rb$. 
$\channel$ is the portion of this overlay between $\lb$ and $\rb$; see Figure~\ref{fig:channel}.
The complexity of $\channel$, denoted by $|\channel|$, is the number of vertices in $\channel$. 

To estimate $\lb(x)$ and $\rb(x)$ , we use 
Manning's equation \cite{manning1890flow}, a widely used empirical formula relating the channel geometry and flow rate as follows.

Let $x$ be a point on an edge $e \in \mesh$ with flow value $\flowrate_e$. 
Let $\ell_x$ be the line in the $xy$-plane passing through $x$ and normal to $e$, 
and let $\Pi = \ell_x \times \reals$ be the vertical plane containing $\ell$. 
Let $\terrain_x = \terrain \cap \Pi$ be the cross-section of $\terrain$ in $\Pi$, which we refer to as the \textit{profile}  of $\terrain$ at $x$. 
$\terrain_x$ is a polygonal chain whose vertices (resp. edges) are the intersection points of edges (resp. faces) 
of $\terrain$ with $\Pi$. 
See Figure~\ref{fig:channel}.
Let $\hat{x} = (x,h(x)$ be the vertex on $\terrain_x$ corresponding to the point $x \in e$, 
i.e. $\hat{x}$ is vertically above $x$. 

We divide $\terrain_x$ into two polygonal rays $L_{x}$, $R_{x}$ at the vertex $\hat{x}$, 
with $L_x$ (resp. $R_x$) lying to the left (resp. right) of $\paths$.  
For a height $z \geq h(x)$, let $\lambda(z)$  (resp. $\rho(z)$) be the first point on $L_{x}$ (resp. $R_{x}$) at height $z$ as we walk along $L_{x}$ (resp. $R_{x}$.) 
Let $A_x(z)$ denote the area of the polygon formed by the segment $\lambda_x(z)\rho_x(z)$ and the portion of 
$\terrain_x$ between $\lambda_x(z)$ and $\rho_{x}(z)$, and let $P_x(z)$ denote the arc length of $\terrain_x$ between $\lambda_x(z)$ and $\rho_x(z)$. 
If the water has height $z$ at $x$, then Manning's equation \cite{manning1890flow} states that the flow rate at $x$ is 
\begin{equation}
    \label{eqn:manning}
   \flowrate_x(z) = \frac{A_x(z)^{5/3}\sigma_e^{1/2}}{\mu_e P_x(z)^{2/3} },
\end{equation}

where 
$\sigma_e$ is the slope in the $z$-direction of the edge of $\terrain$ corresponding to $e$, 
and $\mu_e$ is Manning's roughness coefficient. 
We assume that we are given the value of $\mu_e$, which depends on the material of the terrain at $e$ (e.g. concrete, dirt, light brush, etc.)   
Manning's equation is typically used to compute the flow rate $\flowrate_{x}(z)$ of rivers given a measurement of the river depth and approximate channel geometry. 
Here instead we state an inverse problem: given the flow rate $\flowrate_{x}$ at $x$, determine the depth and width of the river at $x$. 
Let $\Delta(x)$ be the value of $z$ for which $\flowrate_{x}(z) = \flowrate_e$. 
We set $\lb(x) = \lambda(\Delta(x))$ and $\rb(x) = \rho(\Delta(x))$, 
i.e., the \textit{river bank} points on $\terrain$ corresponding to $x$;
Let $\channel_x = \terrain[\lb(x),\rb(x)]$ be the profile of the channel at $x$. 
See Figure \ref{fig:channel}.

We first describe how we compute $\Delta(x)$ for a fixed $x$ and then describe how to track $\lb$ and $\rb$ as we vary $x$. 
For simplicity, we make the following two assumptions:
\begin{itemize}
    \item[(A1)] $\channel_x$ is unimodal for all $x \in I$.
    \item[(A2)] The point $\hat{x}$ is the unique minimum of $\channel_x$. 
\end{itemize}
We discuss in Section~\ref{subsec:general-channels} how to relax these assumptions. 

\subsection{Computing $\Delta(x)$}
Recall that we assume $\channel_x$ to be unimodal with $\hat{x}$ as its unique minimum. 
Without loss of generality, assume that the edge $e$ containing $x$ is parallel to the $x$-axis, 
so $\Pi$ is parallel to the $yz$-plane. 
We raise the value of $z$ starting from $h(x)$ and stopping at the height of vertices of 
$\terrain_x$ until we find a vertex $\hat{v} = (v,h(v))$ such that
$\flowrate_{x}(h(v)) \geq \flowrate_{e}$. 
We now know the edges of $L_x$ and $R_x$ that contain $\lb(x)$ and $\rb(x)$. 
We then compute the points themselves. 

We now describe the procedure in more detail. 
Let $f_{1,x},f_{2,x},\cdots$ be the sequence of edges of $R_x$, 
ordered from left to right, and let $e_{i-1,x},e_{i,x}$ be the endpoints of $f_{i,x}$; $e_{0,x} = x$. 
Recall that each edge $f_{i,x}$ is the intersection of a face $f_i$ of $\terrain$ 
with the vertical plane $\Pi$, and each endpoint $e_{j,x}$ is $e_j \cap \Pi$ for some edge $e_j$ of $\terrain$. 
For each edge $f_{i,x}$, let $f^{\uparrow}_{i,x}$ be the semi-infinite trapezoid 
formed by the edge $f_{i,x}$ and the vertical rays in the $(+z)$-direction from the endpoints $e_{i-1,x},e_{i,x}$ of $f_{i,x}$. 
For a value $z_0 \in \reals$, we define the trapezoid $f^{\uparrow}_{i,x}(z_0)$ to be the intersection of 
$f^{\uparrow}_{i,x}$ with the halfspace $z \leq z_0$; 
$f^{\uparrow}_{i,x}(z_0)$ may be empty, or it may be a triangle. 
We define $A_{i,x}(z)$ to be the area of $f^{\uparrow}_{i,x}(z)$ 
and $P_{i,x}(z)$ to be the length of the bottom edge of $f^{\uparrow}_{i,x}(z)$, which is a portion of the edge $f_{i,x}$.
Let $e_{i,x} = (x,a_{i}(x),b_{i}(x))$ denote the coordinates of $e_{i,x}$ as a function of $x$, 
and set $w_{i}(x) = a_{i}(x)-a_{i-1}(x)$ and $h_{i}(x) = b_{i}(x)-b_{i-1}(x)$. 
Then $A_{i,x}$ can be written as 
\begin{equation} 
    \label{eqn:area}
    A_{i,x}(z) = 
    \begin{cases}
        0 &\text{$z < b_{i-1}(x)$},\\
        \frac{(z-b_{i-1})^2w_i(x)}{2h_i(x)}  &\text{$b_{i-1}(x)$ < z < $b_i(x)$},\\
        w_{i}(x)(\frac{1}{2}h_i(x)+(z-b_i(x)) &\text{$b_i(x) < z$}.
    \end{cases}
\end{equation}
Similarly $P_{i,x}$ can be expressed as
\begin{equation}
    \label{eqn:perim}
    P_{i,x}(z) = 
    \begin{cases}
        0 &\text{$z < b_{i-1}(x)$},\\
        \|f_i(x)\| \frac{(z-b_{i-1})}{h_i(x)} &\text{$b_{i-1}(x) < z < b_i(x)$},\\
        \|f_i(x)\|&\text{$b_i(x) < z$}.
    \end{cases}
\end{equation}

We note that $P_{i,x}$ (resp. $A_{i,x}$) is a piecewise-linear (resp. piecewise-quadratic) function of $z$ for a fixed $x$. 
We can define $F_{j,x}(z), P_{j,x}(z)$ and $A_{j,x}(z)$ for the edges $f_{j,x}$ of $L_x$ as well. 
We can now express $P_x$ and $A_x$ as:

\begin{equation}
    \label{eqn:area-perim-sum}
    P_x(z) = \sum_{i}P_{i,x}(z) \quad\text{and}\quad A_{x}(z) = \sum_{i} A_{i,x}(z),
\end{equation}

where the summation is taken over all edges of $\terrain_x$ that contain a point of height at most $z$. 
$P_x$ and $A_x$ are also piecewise-linear and piecewise-quadratic functions respectively. 

Let $z_0 < z_1 < z_2 < \cdots$ be the heights of vertices of $\terrain_x$. 
$P_x(z_0) = A_x(z_0) = 0$. 
Assuming $P_{i,x}(z_{i-1}), A_{i,x}(z_{i-1})$ have been computed 
then $P_{i,x}(z_i)$, $A_{i,x}(z_i)$ can be computed in $\BigO(1)$ time using (\ref{eqn:area})--(\ref{eqn:area-perim-sum}). 

Let $z_k$ be the first value for which $\flowrate_{x}(z_k) \geq \flowrate_e$. 
Since $P_{i,x}(z)$ (resp. $A_{i,x}(z)$) is a linear (resp. quadratic) function for $z \in (z_{k-1},z_k)$, 
the value of $\Delta(x) \in (z_{k-1},z_k]$ can be computed in $\BigO(1)$ time by plugging these functional forms in (\ref{eqn:manning}). 
Let $f_{L,x}$ (resp. $f_{R,x}$) be the edge of $L_x$ (resp. $R_x$) at which the vertical sweep stopped. 
Then $\lb(x)$ (resp. $\rb(x)$) is the point on $f_{L,x}$ (resp. $f_{R,x}$) of height $\Delta(x)$ 
and can be computed in $\BigO(1)$ time. 
The total time spent by the procedure is $\BigO(|\channel_x|)$. 
Hence, we obtain the following.

\begin{lemma}
    \label{lem:channel} For a given point $x \in \paths$, $\Delta(x)$, $\lb(x)$ and $\rb(x)$ can be computed in $\BigO(|\channel_x|)$ time. 
\end{lemma}

\subsection{Unimodal Channel Algorithm}
We now describe an algorithm for compute the channel $\channel$ assuming (A1) and (A2) hold. 
Recall that for any $x \in I$, the vertices of $\channel_x$ are intersection points of the edges with the plane
$\Pi_x$. 
Let $\Gamma_{x} = \left< \gamma_1, \cdots, \gamma_u \right>$ denote the sequence of these edges, 
which implicitly define $\channel_x$. 
We refer to $\Gamma_x$ as the \textit{combinatorial structure} of $\channel_x$. 
We compute $\channel$ by varying $x$ continuously from $x_0$ to $x_1$ and maintaining $\channel_x$. 
As $x$ varies, $\channel_x$ changes continuously, i.e., each vertex of $\channel_x$ traces a curve, 
but the combinatorial structure $\Gamma_{x}$ changes only at discrete values of $x$, called the \textit{events}. 
The algorithm works by sweeping the line $\ell_x$ along $\paths$, stopping at events as we traverse $\paths$. 
As long as $x$ lies on the same edge of $\paths$, $\ell_x$ simply translates. 
At vertices of $\paths$, where the sweep line $\ell_x$ shifts from one edge to the next one in $\paths$, 
the algorithm continues by rotating $\ell_x$ to make it normal to the next edge. 
We first describe how we sweep along an edge of $\paths$ and then describe how the sweep line rotates at a vertex of $\paths$. 

\header{Edges.}
Fix an edge $e \in \paths$. 
Without loss of generality, assume that $e$ is parameterized as $e:[0,1] \to \reals^2$. 

As we sweep along $e$ and vary $x$, the left and right banks $\lb, \rb$ trace curves that lie inside fixed faces of $\terrain$ and the remaining vertices of $\channel_x$ trace the corresponding edges of $\terrain$. 
The algorithm encounters the following two types of events at which $\Gamma_x$ changes: 
\begin{enumerate}
    \item the sweep line reaches an endpoint of an edge $(u',v')$ of $\Gamma_x$  which is a vertex of $\terrain$, or 
    \item $\lb$ or $\rb$ intersects an edge $e'$ (bounding the face containing it) of $\terrain$. 
\end{enumerate}
The first event results in one or more edges in $\channel_x$ shrinking to the point $v'$, 
and one or more new edges starting at $v'$. 
The second event results in either the addition and/or removal of an extremal edge in $\channel_x$ 
(and thus insertion/deletion of an edge in $\Gamma_x$) depending on whether the height of the channel is increasing or decreasing. 

The first type of events are easy to detect, as they correspond to the vertices of $\terrain$. 
The second type are more challenging, and we detect them as follows. 
By maintaining functions representing the area and wetted perimeter of $\channel_x$ 
as a bivariate function of $x$ and $z$ (using (\ref{eqn:area}-\ref{eqn:area-perim-sum}))
and using Manning's equation we can express $\Delta(x)$ as a function of $x$. 
We then compute when $\Delta(x)$ reaches the top or bottom boundary of the face containing $\lb$ (resp. $\rb$). 
These time instances correspond to the second type of event. 

Process the events in order, by popping the first event from the priority queue $Q$. 
If it is the first type of event corresponding to a vertex $v$ of $\terrain$, 
we remove from $\Gamma$ the edges that end at $v$ and insert into $\Gamma$ the edges that start at $v$. 
We add the other endpoints of the new edges to $Q$ as new first type of events. 
We also remove from $A(x,z)$ and $P(x,z)$ the terms corresponding to the old edges 
and add terms corresponding to the new edges. 
If an edge of the face of $\terrain$ containing $\lb$ or $\rb$ changes, 
we also update the second type of event in $Q$. 

If the event corresponds to $\lb$ or $\rb$ reaching an edge of the terrain, 
either add the new face the it crosses into and/or remove the face it crosses out of.
We update $\Gamma$ as well as $Q$. 
We also $A(z)$ and $P(z)$ accordingly. 

$A(x,z)$ and $P(x,z)$ can be maintained using a height balanced tree so that its functional form can be updated in 
$\BigO(\log n)$ time per insertion/deletion of a term in $A(x,z)$ and $P(x,z)$. 

We continue this process until we reach the event corresponding to $x=1$, when the endpoint of $e$ is reached. 
Let $|\channel_e|$ be the number of total faces contained in the channel from $\channel(0)$ to $\channel(1)$. 
Between any two events, $\lb(x)$ and $\rb(x)$  intersect a face only a constant number of times.
Therefore the total number of events is $\BigO(|\channel_e|)$, giving a total running time to sweep an edge of $O( |\channel_e| \log |\channel_e|)$. 

\begin{figure}[t]%
  \centering
  \includegraphics[]{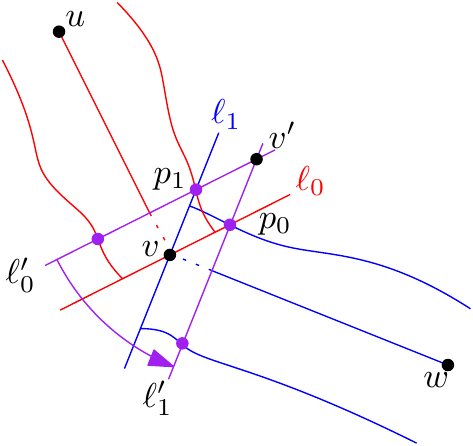}%
  \caption{%
    \normalfont
    The river profile is taken as the union of the profile of the two edges and that as we rotate around the vertex from $\ell_0$ to $\ell_1$.
  }
  \label{fig:vert-rot}%
\end{figure}

\header{Vertex.}
When transitioning from one edge to another along the path, we must join the two channels. 
We will assume the there are no sharp bends between two edges of $\paths$, 
specifically the angle between the sweep line along the first and second channel is less than 90 degrees, 
Taking the two channels $\channel_{(u,v)}$ and $\channel(v,w)$ we see that on one side of $v$ the two channels will intersect, while on the other they will be disjoint. 
Let $\ell_0$ (resp. $\ell_1$) be the sweep line perpendicular to $(u,v)$ (resp. $(v,w)$) at $v$, 
and $p_0$ (resp. $p_1$) be the intersection point between $\ell_0$ with the riverbank of $\channel_{(v,w)}$ 
(resp. $\ell_1$ with the riverbank of $\channel_{(u,v)}$.) 
Then let $\ell_0'$ (resp. $\ell_1'$) be the line parallel to $\ell_0$ at $p_1$ (resp. $\ell_1$ at $p_0$), 
and $v'$ be the intersection point between $\ell_0'$ and $\ell_1'$. 
Now rotating about $v'$ from $\ell_0'$ to $\ell_1'$ will sweep the area where the two channels intersect. 
See Figure \ref{fig:vert-rot}. 

In a similar manner in which we swept along each edge, 
let $\channel_{v',\theta}$ be the profile of $v'$ with a line $\ell$ at angle $\theta$. 
As we rotate the sweep line, we assume the water is flowing at the intersection of $\ell$ with either $(u,v)$ or $(v,w)$.
We will similarly maintain the area function $A(\theta,z)$ (resp. the wetted perimeter function $P(\theta,z)$) as a 
sum of functions on each face in the channel. 
Additionally when computing Manning's equation, as we rotate we will linearly interpolate between slope values 
as well as the roughness coefficients for the two edges. 
The main difference is now for each face $f_i \in \faces$, $h_i(\theta,z)$ and $w_i(\theta,z)$ are not linear functions in $\theta$ as it was when we swept along an edge. 
However, we can still computed analytically the events corresponding to the riverbank(s) crossing edges of faces. 

Letting $|\channel_v|$ denote the total number of faces in the channel obtained by sweeping at $v$. 
Then the total time spent is $\BigO|\channel_v| \log|\channel_v|)$

Putting everything together we obtain the following:

\begin{theorem}
    \label{thm:2dchannel} 
    Given a triangulation $\mesh$ with $n$ vertices, a height function $h: \mesh \to \reals$ which is linear on each face of $\mesh$ a path $\paths$ in $\mesh$, if the channel $\channel_x$ is unimodal for all $x$ along the path, we can compute the 2D flow network in time $\BigO(|\channel| \log |\channel|)$ where $|\channel|$ is the total number of faces in the channel obtained by sweeping along the edges and vertices of $\paths$. 
\end{theorem}

\subsection{General Channels}
\label{subsec:general-channels}
We will now show how to modify the algorithm when assumptions (A1) and (A2) do not hold.
When (A2) does not hold, i.e. $\hat{x}$ is not a local minima, the modification is straightforward. 
Simply walk down the edges of $\Gamma_x$ until finding a local minima $x'$ and then run the algorithm using that vertex as the division point for the polygonal rays $L_{x'}$ and $R_{x'}$. 
It may be the case that the edge containing $x'$ ends and a new minima $x''$ begins, but this can be handled like any other event in the algorithm. 
The only difference comes in the interpretation and runtime analysis. 
When (A1) does not hold, it may be the case that the water level in the 2D channel does reach an edge along which water was flowing according to the 1D flow network. 
However this is reasonable behavior, as the 1D flow network assumes water only flows along edges of the terrain whereas in reality water will also flow along the faces. 
For the runtime analysis, we now include the edges searched along from $x$ to $x'$ to be included in the channel $\channel_e$. 

\begin{figure}[t]%
  \centering
  \includegraphics[]{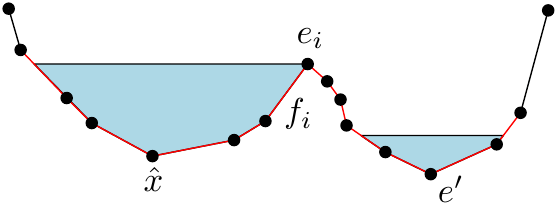}%
  \caption{%
    \normalfont A channel profile $\terrain_x$ with local maxima $y$. 
    If the flow in the primary channel on the right causes the channel height to exceed $h(y)$, 
    excess spills to the secondary channel on the left. 
  }
  \label{fig:nonconvex}%
\end{figure}

When (A1) does not hold, i.e. the channel is not unimodal, 
when $\lb$ or $\rb$ intersect an edge of $\terrain$ bounding the face containing it, 
that edge may correspond to a local maximum in the profile. 
In this case, water flows over the ridge into a secondary channel. 
See Figure \ref{fig:nonconvex}\footnote{Here $|\channel_x|$ contains all faces between the leftmost and rightmost river banks, i.e., all edges marked red.}. 
We must now account for the decrease of the flow-rate in the primary channel as well as determining the height of water in the secondary channel.

\header{Updating flow rates.}
Let $f_i$ be the first face containing a local maximum encountered when sweeping along an edge maintaining $\Gamma_x$, and let $e_{i}$ be the edge of $\terrain$ corresponding to the maximum. 
Using Manning's equation, we compute the flow rate corresponding to when this unimodal channel becomes full $\flowrate_{f_{i,x}} = \flowrate_{x}(h(e_{i,x})$.
If $\flowrate_{e} > \flowrate_{f_{i,x}}$ a spill will occur, and the excess flow is sent to the local minimum in the secondary channel. 
To account for the water flowing out of the channel, consider the flow rate along $e$ as a function of $x$, $\flowrate_{e}(x)$. 
Letting $e'$ be the minimum edge in the secondary channel, we set $\flowrate_{e'}(x) = \flowrate_{e}-\flowrate_{f_i}(x)$. 
As we continue the sweep $\flowrate_{e}(x)$ will be a non-increasing function, we set $\flowrate_{e}(x) = \min (\flowrate_e,\flowrate_{f_{i,x}}, \min_{x' < x} \flowrate_{e}(x'))$. 

\header{Secondary channel.}
When an overflow occurs we continue adding faces to the profile until finding the next local minima corresponding to edge $e'$. 
Then in this secondary channel if the excess flow exceeds our threshold for the 1D channel, i.e. $\flowrate_{e'}(x) \geq \psi$, 
we repeat the unimodal channel algorithm in this channel using $\flowrate_{e'}(x)$. 
It may be the case that $\flowrate_{e'}(x)$ is enough to fill the secondary channel above the height of a ridge. 
If it is the same ridge that spilled into the secondary channel we treat it as a single channel. 
If it is a ridge further out, we repeat the process in the tertiary channel.

\section{Conclusion}
In this paper we presented algorithms for a number of flow routing problems:
First, we developed a fast internal-memory algorithm for the terrain-flow query problem which given a rain distribution, computes the flow rate $\flowrate_{e}$ for all edges $e \in \terrain$.
We also showed how the algorithm could be made I/O-efficient.
Next, we presented a faster algorithm if one is interested in computing the flow rate of only one edge, after some preprocessing.
Finally, given a flow path along the edges of $\terrain$, we proposed an algorithm to determine the 2D channel along which water flows; our algorithm does not make any assumption about the geometry of the channel.

We conclude by mentioning a few directions for future work.
The model of extracting 2D channels leaves a number of open questions. 
For instance, if the 1D flow network is a forest then channels along different
paths will interact which leaves the question of how we merge these channels to
construct a 2D river network.

While we consider the flow rate as a function of time, it only changes when the rain distribution changes or a spill event occurs. 
That is, the effects of such events are propagated to all reachable vertices instantaneously. 
While this assumption is reasonable for local effects and for flash floods when a large volume of rain falls over a short duration, an interesting question is to make the model more general and account for the time it takes water to flow over the terrain.


\balance
\bibliography{references}


\begin{thebibliography}{17}


\ifx \showCODEN    \undefined \def \showCODEN     #1{\unskip}     \fi
\ifx \showDOI      \undefined \def \showDOI       #1{#1}\fi
\ifx \showISBNx    \undefined \def \showISBNx     #1{\unskip}     \fi
\ifx \showISBNxiii \undefined \def \showISBNxiii  #1{\unskip}     \fi
\ifx \showISSN     \undefined \def \showISSN      #1{\unskip}     \fi
\ifx \showLCCN     \undefined \def \showLCCN      #1{\unskip}     \fi
\ifx \shownote     \undefined \def \shownote      #1{#1}          \fi
\ifx \showarticletitle \undefined \def \showarticletitle #1{#1}   \fi
\ifx \showURL      \undefined \def \showURL       {\relax}        \fi
\providecommand\bibfield[2]{#2}
\providecommand\bibinfo[2]{#2}
\providecommand\natexlab[1]{#1}
\providecommand\showeprint[2][]{arXiv:#2}

\bibitem[\protect\citeauthoryear{Aggarwal and Vitter}{Aggarwal and
  Vitter}{1988}]%
        {AV88}
\bibfield{author}{\bibinfo{person}{A Aggarwal} {and} \bibinfo{person}{JS
  Vitter}.} \bibinfo{year}{1988}\natexlab{}.
\newblock \showarticletitle{The input/output complexity of sorting and related
  problems}.
\newblock \bibinfo{journal}{\emph{Commun. ACM}} \bibinfo{volume}{31},
  \bibinfo{number}{9} (\bibinfo{year}{1988}), \bibinfo{pages}{1116--1127}.
\newblock


\bibitem[\protect\citeauthoryear{Arge, Rav, Raza, and Revsb{\ae}k}{Arge
  et~al\mbox{.}}{2017}]%
        {ARRR17}
\bibfield{author}{\bibinfo{person}{L Arge}, \bibinfo{person}{M Rav},
  \bibinfo{person}{S Raza}, {and} \bibinfo{person}{M Revsb{\ae}k}.}
  \bibinfo{year}{2017}\natexlab{}.
\newblock \showarticletitle{I/O-Efficient Event Based Depression Flood Risk}.
  In \bibinfo{booktitle}{\emph{Proc. 19th Workshop on Algorithm Engineering and
  Experiments}}. \bibinfo{pages}{259--269}.
\newblock


\bibitem[\protect\citeauthoryear{Arge and Revsb{\ae}k}{Arge and
  Revsb{\ae}k}{2009}]%
        {AR09}
\bibfield{author}{\bibinfo{person}{L Arge} {and} \bibinfo{person}{M
  Revsb{\ae}k}.} \bibinfo{year}{2009}\natexlab{}.
\newblock \showarticletitle{{I/O}-efficient Contour Tree Simplification}. In
  \bibinfo{booktitle}{\emph{Intl. Sympos. on Algos. and Computation}}.
  \bibinfo{pages}{1155--1165}.
\newblock


\bibitem[\protect\citeauthoryear{Arge, Revsb{\ae}k, and Zeh}{Arge
  et~al\mbox{.}}{2010}]%
        {ARZ10}
\bibfield{author}{\bibinfo{person}{L Arge}, \bibinfo{person}{M Revsb{\ae}k},
  {and} \bibinfo{person}{N Zeh}.} \bibinfo{year}{2010}\natexlab{}.
\newblock \showarticletitle{{I/O}-efficient computation of water flow across a
  terrain}. In \bibinfo{booktitle}{\emph{Proc. 26th Annu. Sympos. on Comp.
  Geom.}} \bibinfo{pages}{403--412}.
\newblock


\bibitem[\protect\citeauthoryear{Arge, Toma, and Vitter}{Arge
  et~al\mbox{.}}{2000}]%
        {arge2000grid}
\bibfield{author}{\bibinfo{person}{L Arge}, \bibinfo{person}{L Toma}, {and}
  \bibinfo{person}{J Vitter}.} \bibinfo{year}{2000}\natexlab{}.
\newblock \showarticletitle{I/O-Efficient Algorithms for Problems on Grid-Based
  Terrains}.
\newblock \bibinfo{journal}{\emph{Journal of Experimental Algorithmics}}
  \bibinfo{volume}{6} (\bibinfo{date}{10} \bibinfo{year}{2000}).
\newblock
\urldef\tempurl%
\url{https://doi.org/10.1145/945394.945395}
\showDOI{\tempurl}


\bibitem[\protect\citeauthoryear{Bates and De~Roo}{Bates and De~Roo}{2000}]%
        {bates2000simple}
\bibfield{author}{\bibinfo{person}{PD Bates} {and} \bibinfo{person}{APJ
  De~Roo}.} \bibinfo{year}{2000}\natexlab{}.
\newblock \showarticletitle{A simple raster-based model for flood inundation
  simulation}.
\newblock \bibinfo{journal}{\emph{Journal of hydrology}} \bibinfo{volume}{236},
  \bibinfo{number}{1-2} (\bibinfo{year}{2000}), \bibinfo{pages}{54--77}.
\newblock


\bibitem[\protect\citeauthoryear{Carr, Snoeyink, and Axen}{Carr
  et~al\mbox{.}}{2003}]%
        {CSA03}
\bibfield{author}{\bibinfo{person}{H Carr}, \bibinfo{person}{J Snoeyink}, {and}
  \bibinfo{person}{U Axen}.} \bibinfo{year}{2003}\natexlab{}.
\newblock \showarticletitle{Computing contour trees in all dimensions}.
\newblock \bibinfo{journal}{\emph{Comp. Geom.}} \bibinfo{volume}{24},
  \bibinfo{number}{2} (\bibinfo{year}{2003}), \bibinfo{pages}{75--94}.
\newblock


\bibitem[\protect\citeauthoryear{Chiang, Goodrich, Grove, Tamassia, Vengroff,
  and Vitter}{Chiang et~al\mbox{.}}{1995}]%
        {CGGTVV95}
\bibfield{author}{\bibinfo{person}{Y Chiang}, \bibinfo{person}{MT Goodrich},
  \bibinfo{person}{EF Grove}, \bibinfo{person}{R Tamassia}, \bibinfo{person}{DE
  Vengroff}, {and} \bibinfo{person}{JS Vitter}.}
  \bibinfo{year}{1995}\natexlab{}.
\newblock \showarticletitle{External-memory graph algorithms}. In
  \bibinfo{booktitle}{\emph{Proc. Sixth Annu. ACM-SIAM Sympos. on Discrete
  Algos.}} \bibinfo{pages}{139--149}.
\newblock


\bibitem[\protect\citeauthoryear{Edelsbrunner, Harer, and
  Zomorodian}{Edelsbrunner et~al\mbox{.}}{2001}]%
        {EHZ01}
\bibfield{author}{\bibinfo{person}{H Edelsbrunner}, \bibinfo{person}{J Harer},
  {and} \bibinfo{person}{A Zomorodian}.} \bibinfo{year}{2001}\natexlab{}.
\newblock \showarticletitle{Hierarchical Morse complexes for piecewise linear
  2-manifolds}. In \bibinfo{booktitle}{\emph{Proc. 17th Annu. Sympos. Comp.
  Geom.}} \bibinfo{pages}{70--79}.
\newblock


\bibitem[\protect\citeauthoryear{Holm, Rotenberg, and Thorup}{Holm
  et~al\mbox{.}}{2015}]%
        {holm2015planar}
\bibfield{author}{\bibinfo{person}{J Holm}, \bibinfo{person}{E Rotenberg},
  {and} \bibinfo{person}{M Thorup}.} \bibinfo{year}{2015}\natexlab{}.
\newblock \showarticletitle{Planar reachability in linear space and constant
  time}. In \bibinfo{booktitle}{\emph{2015 IEEE 56th Annual Symposium on
  Foundations of Computer Science}}. IEEE, \bibinfo{pages}{370--389}.
\newblock


\bibitem[\protect\citeauthoryear{Kreveld, Oostrum, Bajaj, Pascucci, and
  Schikore}{Kreveld et~al\mbox{.}}{1997}]%
        {KOBPS97}
\bibfield{author}{\bibinfo{person}{M Kreveld}, \bibinfo{person}{R Oostrum},
  \bibinfo{person}{C Bajaj}, \bibinfo{person}{V Pascucci}, {and}
  \bibinfo{person}{D Schikore}.} \bibinfo{year}{1997}\natexlab{}.
\newblock \showarticletitle{Contour trees and small seed sets for isosurface
  traversal}. In \bibinfo{booktitle}{\emph{Proc. 13th Annu. Sympos. on Comp.
  Geom.}} \bibinfo{pages}{212--220}.
\newblock


\bibitem[\protect\citeauthoryear{Liu and Snoeyink}{Liu and Snoeyink}{2005}]%
        {LS05}
\bibfield{author}{\bibinfo{person}{Y Liu} {and} \bibinfo{person}{J Snoeyink}.}
  \bibinfo{year}{2005}\natexlab{}.
\newblock \showarticletitle{Flooding triangulated terrain}. In
  \bibinfo{booktitle}{\emph{Proc. 11th Intl. Sympos. on Spatial Data
  Handling}}. \bibinfo{pages}{137--148}.
\newblock


\bibitem[\protect\citeauthoryear{Lowe and Agarwal}{Lowe and Agarwal}{2019}]%
        {lowe2019flood}
\bibfield{author}{\bibinfo{person}{A Lowe} {and} \bibinfo{person}{PK Agarwal}.}
  \bibinfo{year}{2019}\natexlab{}.
\newblock \showarticletitle{Flood-Risk Analysis on Terrains under the
  Multiflow-Direction Model}.
\newblock \bibinfo{journal}{\emph{ACM Trans. Spatial Algorithms Syst.}}
  \bibinfo{volume}{5}, \bibinfo{number}{4}, Article \bibinfo{articleno}{26}
  (\bibinfo{date}{Sept.} \bibinfo{year}{2019}), \bibinfo{numpages}{27}~pages.
\newblock
\showISSN{2374-0353}
\urldef\tempurl%
\url{https://doi.org/10.1145/3340707}
\showDOI{\tempurl}


\bibitem[\protect\citeauthoryear{Manning, Griffith, Pigot, and
  Vernon-Harcourt}{Manning et~al\mbox{.}}{1890}]%
        {manning1890flow}
\bibfield{author}{\bibinfo{person}{R Manning}, \bibinfo{person}{JP Griffith},
  \bibinfo{person}{TF Pigot}, {and} \bibinfo{person}{LF Vernon-Harcourt}.}
  \bibinfo{year}{1890}\natexlab{}.
\newblock \bibinfo{booktitle}{\emph{On the flow of water in open channels and
  pipes}}.
\newblock


\bibitem[\protect\citeauthoryear{Rav, Lowe, and Agarwal}{Rav
  et~al\mbox{.}}{2017}]%
        {RLA17}
\bibfield{author}{\bibinfo{person}{M Rav}, \bibinfo{person}{A Lowe}, {and}
  \bibinfo{person}{PK Agarwal}.} \bibinfo{year}{2017}\natexlab{}.
\newblock \showarticletitle{Flood Risk Analysis on Terrains}. In
  \bibinfo{booktitle}{\emph{Proc. of the 25th ACM SIGSPATIAL Int. Conference on
  Advances in GIS}}. ACM, \bibinfo{pages}{36}.
\newblock


\bibitem[\protect\citeauthoryear{Sanders}{Sanders}{2001}]%
        {sanders2000queue}
\bibfield{author}{\bibinfo{person}{Peter Sanders}.}
  \bibinfo{year}{2001}\natexlab{}.
\newblock \showarticletitle{Fast Priority Queues for Cached Memory}.
\newblock \bibinfo{journal}{\emph{ACM J. Exp. Algorithmics}}
  \bibinfo{volume}{5} (\bibinfo{date}{Dec.} \bibinfo{year}{2001}),
  \bibinfo{pages}{7–es}.
\newblock
\showISSN{1084-6654}
\urldef\tempurl%
\url{https://doi.org/10.1145/351827.384249}
\showDOI{\tempurl}


\bibitem[\protect\citeauthoryear{Wood, Neal, Bates, Hostache, Wagener,
  Giustarini, Chini, Corato, and Matgen}{Wood et~al\mbox{.}}{2016}]%
        {wood2016calibration}
\bibfield{author}{\bibinfo{person}{M Wood}, \bibinfo{person}{JC Neal},
  \bibinfo{person}{PD Bates}, \bibinfo{person}{R Hostache}, \bibinfo{person}{T
  Wagener}, \bibinfo{person}{L Giustarini}, \bibinfo{person}{M Chini},
  \bibinfo{person}{G Corato}, {and} \bibinfo{person}{P Matgen}.}
  \bibinfo{year}{2016}\natexlab{}.
\newblock \showarticletitle{Calibration of channel depth and friction
  parameters in the LISFLOOD-FP hydraulic model using medium resolution SAR
  data and identifiability techniques}.
\newblock \bibinfo{journal}{\emph{Hydrology and Earth System Sciences}}
  \bibinfo{volume}{20}, \bibinfo{number}{12} (\bibinfo{year}{2016}),
  \bibinfo{pages}{4983--4997}.
\newblock


\end{thebibliography}
\end{document}